\PassOptionsToPackage{table,xcdraw}{xcolor}
\documentclass[conference]{IEEEtran}
\IEEEoverridecommandlockouts
\usepackage{booktabs}   %% For formal tables:
\usepackage{subcaption} %% For complex figures with subfigures/subcaptions
\usepackage{array}
\usepackage{amsmath,amssymb,amsfonts}
\usepackage{algorithm}
\usepackage[noend]{algpseudocode}
\usepackage{graphicx}
\usepackage{textcomp}
\usepackage{float}
\usepackage{listings}
\usepackage{xspace}
\usepackage{multirow}
\usepackage{amsthm}

\usepackage[skins]{tcolorbox}

\usepackage [latin1]{inputenc}
\usepackage{balance}

\newtcolorbox{myframe}[2][]{%
  enhanced,colback=white,colframe=black,coltitle=black,
  sharp corners,
  toprule=1.0pt,
  rightrule=0.3pt,
  leftrule=0pt,
  bottomrule=0pt,
  fonttitle=\itshape\scshape\large,
  left=0pt,right=5pt,top=5pt,bottom=3pt,
  attach boxed title to top right={yshift=-0.3\baselineskip-0.4pt,xshift=-5mm},
  boxed title style={tile,size=minimal,left=0.2mm,right=0.5mm,
    colback=white,before upper=\strut},
  title=#2,#1
}

\newcommand{\tool}{\textsc{DeepName}\xspace}

\newtheorem{Definition}{Definition}

\newcolumntype{L}[1]{>{\raggedright\arraybackslash}p{#1}}

\newcommand{\code}[1]{{\footnotesize\textsf{#1}}}
\usepackage{amsthm}
 \definecolor{dkgreen}{rgb}{0,0.6,0}
\definecolor{gray}{rgb}{0.5,0.5,0.5}
\definecolor{mauve}{rgb}{0.58,0,0.82}
\begin{document}

\title{A Context-based Automated Approach for Method Name Consistency Checking and Suggestion}

\author{\IEEEauthorblockN{Yi Li}
\IEEEauthorblockA{\textit{Department of Informatics} \\
  \textit{New Jersey Institute of Technology}\\
  New Jersey, USA \\
Email: yl622@njit.edu}
\and
\IEEEauthorblockN{Shaohua Wang\IEEEauthorrefmark{1}}
\IEEEauthorblockA{\textit{Department of Informatics} \\
  \textit{New Jersey Institute of Technology}\\
  New Jersey, USA \\
Email: davidsw@njit.edu}
\and
\IEEEauthorblockN{Tien N. Nguyen}
\IEEEauthorblockA{\textit{Computer Science Department} \\
  \textit{The University of Texas at Dallas}\\
  Texas, USA \\
Email: tien.n.nguyen@utdallas.edu}
}

\maketitle
\begingroup\renewcommand\thefootnote{\IEEEauthorrefmark{1}}
\footnotetext{Corresponding Author}
\endgroup
\begin{abstract}
Misleading method names in software projects can confuse developers,
which may lead to software defects and affect code understandability.
In this paper, we present {\tool}, a context-based, deep learning
approach to detect method name inconsistencies and suggest a proper
name for a method. The key departure point is the philosophy of {\em
``Show Me Your Friends, I'll Tell You Who You Are''}.  Unlike the
state-of-the-art approaches, in addition to the method's body, we also
consider the {\em interactions} of the current method under study with
the other ones including the caller and callee methods, and the
sibling methods in the same enclosing class.  The sequences of
sub-tokens in the program entities' names in the contexts are
extracted and used as the input for an RNN-based encoder-decoder to
produce the representations for the current method.  We modify that
RNN model to integrate the copy mechanism and our newly developed
component, called the {\em non-copy mechanism}, to emphasize on the
possibility of a certain sub-token {\em not to be copied to follow}
the current sub-token in the currently generated method name.

We conducted several experiments to evaluate {\tool} on large datasets
with +14M methods. For consistency checking, {\tool} improves the
state-of-the-art approach by 2.1\%, 19.6\%, and 11.9\% relatively in
recall, precision, and F-score, respectively. For name suggestion,
{\tool} improves relatively over the state-of-the-art approaches in
precision (1.8\%--30.5\%), recall (8.8\%--46.1\%), and F-score
(5.2\%--38.2\%). To assess {\tool}'s usefulness, we detected
inconsistent methods and suggested new method names in active
projects.  Among 50 pull requests, 12 were merged into the main
branch. In total, in 30/50 cases, the team members agree that our
suggested method names are more meaningful than the current names.
\end{abstract}

\begin{IEEEkeywords}
Naturalness of Software, Deep Learning, Entity Name
Suggestion, Inconsistent Method Name Checking.
\end{IEEEkeywords}

%, Context-based Name Suggestion.

\section{Introduction}

%An important software quality is source code understandability.
Meaningful and succinct names of program entities play a vital role in
code understandability~\cite{Allamanis_15}. Misleading names in
software libraries can confuse developers and cause them make API
misuses, leading to serious defects~\cite{naming_bug}. During~software
development, the name of a method can become inconsistent with respect
to its intended functionality. The first scenario is that the
inconsistency occurs during the coding of a method, when a misleading
or confusing name is given to the method. In the second scenario,
%despite that the name might be consistent at first, however
inconsistency occurs during software evolution in which code
changes make the method's name and its implementation become
inconsistent with one another.

Several approaches were proposed to detect the
inconsistency between the methods' names and source code, and to
suggest an alternative name if such inconsistency occurs.  The
approaches follow mainly two directions:~{\em information retrieval}
(IR)~\cite{jiang-ase19,kim-icse19} and {\em machine learning}
(ML)~\cite{Allamanis_15,Allamanis_16,alon_popl19,icse20-methodname}.
The idea of IR approaches is that similar methods should have similar
names~\cite{kim-icse19}. Thus, they search~for the names of methods
with similar bodies to suggest for a method with an inconsistent name.
The IR~approaches~generally follow a searching strategy, thus, cannot
recommend~a~new method name that is un-seen in the training data.  The
second direction is machine learning
(ML)~\cite{Allamanis_15,Allamanis_16,alon_popl19,icse20-methodname}.
The ML approaches can overcome the key limitation of the IR direction
due to its capability of generating a new name.  While {\em
  code2vec}~\cite{alon_popl19} generates the method's name based on
the paths over the AST of its body,
%the abstract syntax tree path-based contexts within its body,
MNire~\cite{icse20-methodname} uses the sub-tokens in the program
entities' names.  Other approaches treat the method name suggestion
problem as the extreme~sum\-marization~\cite{Allamanis_16} from the
method's body into a short text. Despite their successes, the
state-of-the-art ML approaches have limitations in dealing with the
methods having little content or the entities' names that are
irrelevant to the functionality.

In this paper, in addition to using the body and interface of the
method under study, we also leverage a philosophy for this naming
problem: {\em ``Show Me Your Friends, I'll Tell You Who You Are''}.
That is, to characterize an entity/person, in addition to using
its/his/her own properties, one can rely on the interactions of that
entity/person with the surrounding and neighboring entities/persons.
For the method name suggestion, examining only the content of the
current method might be insufficient.
The surrounding and interacting methods of a method $m$ under study
could include the methods that are called within the body of $m$ ({\em
  callees}), and the methods that are calling $m$ ({\em callers}).
The neighboring methods are the ones within the same class with $m$
({\em siblings}). The information from the {\em enclosing class} also
provides features for such characterization of a method. The key
features from the {\em caller} and {\em callee} methods, {\em sibling}
methods, and the {\em enclosing class} are used in addition to the
features from the {\em internal body} and {\em interface} of the
method $m$ to verify the consistency of the method with regard to its
name, and to suggest a proper method name. Each of those sources
constitutes a context that is helpful for method name consistency
checking and recommendation. Some methods have little content, but
with sufficient contexts and vice versa. Thus, all contexts are
complementary to one another in name consistency checking and
suggestion.

%Some methods have little internal context, but with sufficient
%interaction context and vice versa. ***However, the cases where all
%contexts have little content are very rare***. Thus, all contexts are
%complementary to one another in helping name suggestion.

We develop {\tool}, a context-based approach for method name
consistency checking and suggestion. For the method $m$ under study,
it extracts the features from four~contexts: the {\em internal context}, the {\em
  caller and callee contexts}, {\em sibling context}, and {\em
  enclosing context}.
For name suggestion, only the callee methods are used because the
callers of $m$ might not be written yet (the current method does not
have a name yet).
We use the name features, specifically, the sequences of sub-tokens
from the program entities within each context, instead of AST or PDG.
It is reported that to infer a method name, using sub-tokens yields
better accuracy than using the AST and PDG of the
method~\cite{icse20-methodname}. The insight is that the naturalness
of names plays crucial role in method name~inference, i.e., method
name depends more on entities' names than AST or PDG (with
data/control flows)~\cite{icse20-methodname}. AST and PDG capture the
structure and procedure of the task, while the method's name is the
summary of the task.
%
%The trivial names/sub-tokens with a single character are removed.
{\tool} uses an RNN-based encoder-decoder to combine all the sequences
of sub-tokens in the contexts into a sequence of
vectors for $m$.
%, which has the integration of all contextual information.
A convolution layer is used on the vector for $m$ to classify
the given name to be consistent or not. To suggest a name for
the given method, we use our vocabulary to map all the generated
vectors into the sub-tokens to compose the method~name.

%In {\tool},

We also modify the operations of the aforementioned
RNN-based encoder-decoder to integrate the {\em copy}
mechanism~\cite{copynet16} and the novel {\em non-copy mechanism}.
%, which is the novel component we design in this work.
A recent study on the methods'
names~\cite{icse20-methodname} has reported that the high
percentage of the sub-tokens of a method name appears in a set of the
sub-tokens from entities' names in a method. Due to this, the copy
mechanism helps emphasize on the possibility of copying certain
sub-tokens from the contexts to the output, i.e, the suggested method
name. The non-copy mechanism is designed to determine the possibility
of a sub-token that {\em must not be copied to follow} the current sub-token in the
currently generated method name.
% Tien
The non-copy mechanism complements to the copy one in the way that it
{\em pushes down the unlikely candidates} (with the sub-tokens not
following a certain token) in the resulting ranked list. Thus, the
likely candidates are pushed up in the list, improving
suggestion~accuracy.
%(Fig.~\ref{toplist}).

%Tien

%Both copy and non-copy mechanisms help {\tool} improve accuracy.

%to leverage the result from Nguyen {\em et
%  al.}~\cite{icse20-methodname} to generate more accurate method
%names.

We conducted several experiments to evaluate {\tool} in method name
consistency checking and in method name recommending on two large
datasets used in prior works with +2M and +14M
methods~\cite{icse20-methodname}.
For inconsistency checking, {\tool} outperformed the
state-of-the-art approaches in Liu {\em et al.}~\cite{kim-icse19} and
MNire~\cite{icse20-methodname} by relatively 13.3\% and 2.1\% in
recall, 34.9\% and 19.6\% in precision, and 25.4\% and 11.9\% in
F-score.
For method name suggestion, {\tool} improves relatively over the
state-of-the-art approaches in both recall (8.8\%--46.1\%) and
precision (1.8\%--30.5\%).  There are 44.3\% of the cases suggested by
{\tool} that exactly match with the correct method names in the
oracle, and 4.7\% of those cases (i.e., 2.1\% of total cases) do not
appear in training data.  This shows that {\tool} can learn to suggest the
method names, rather than retrieving what have been stored. In total,
there are 11.9\% of the cases in which the names are not previously
seen in the training data. The precision and recall of this set of
generated names are 57.6\% and 55.1\% respectively.  To assess
usefulness, we made 50 pull requests (PRs) on the suggesting new names for
the inconsistent methods as detected by {\tool}. Among them, 12 PRs
were actually merged into the main branch, and 18 were approved for
later merging. In total, there are 60\% of the cases that team members
agreed that our suggested names are more meaningful than the current
names. This paper makes the following contributions:

%{\bf A. Empirical Study:} Our results confirm and provide empirical
%evidence for the principle of naturalness of
%software~\cite{hindle-icse12} on the regularity at the token level of
%program entities' names.

%Tien
{\bf A. Representation and Tool:} A novel approach that uses both
internal and interaction contexts for method name consistency checking
and suggestion.

%to detect method name inconsistencies and recommend method names,
%leveraging contextual information with the philosophy of {\em ``Show
%  Me Your Friends, I'll Tell You Who You Are''}.

{\bf B. Novel technique:} In {\tool},
%instead of directly using a deep learning model,
we modify an RNN-based encoder-decoder to integrate a newly developed
mechanism, called {\em Non-copy} mechanism, to help our model pushes
correct candidates to the top, improving top-ranked accuracy.

%select better the next sub-tokens in a suggested method~name.

{\bf C. Empirical Results:} Our empirical evaluation shows
that 1) {\tool} is useful and more accurate than the state-of-the-art
approaches in real-world projects and in a live study;
%Tien
%in method names inconsistency detection and suggestion for real-world
%projects.
%
%We show that it outperforms the state-of-the-art approaches in both
%inconsistency detection and method name suggestion.
%2) as a surprising finding that for method name suggestion, relying on
%the regularity of the tokens of the entities' names in the context
%yields better results than using the code structures (AST) and
%dependencies (PDG). For detailed results, see our
%website~\cite{website}.
2) all four contexts complement to one another and contribute much to
high accuracy. Our replication package is in~\cite{DeepName2021}.

\section{Motivating Examples}
\label{motiv:sec}

\subsection{Examples}
\label{motivexample}

%In this section, we will present a few examples and our observations to motivate our approach.

\begin{figure}[t]
	\centering
	\renewcommand{\lstlistingname}{Method}
	\lstset{
		numbers=left,
		numberstyle= \tiny,
		keywordstyle= \color{blue!70},
		commentstyle= \color{red!50!green!50!blue!50},
		frame=shadowbox,
		rulesepcolor= \color{red!20!green!20!blue!20} ,
%                xleftmargin=1.5em,xrightmargin=1em, aboveskip=1em,
%		framexleftmargin=0.5em,
%                numbersep= 0 em,
%		language=Java,
%		basicstyle=\scriptsize\ttfamily,
                xleftmargin=1.5em,xrightmargin=0em, aboveskip=1em,
		framexleftmargin=1.5em,
                numbersep= 5pt,
		language=Java,
                basicstyle=\scriptsize\ttfamily,
                numberstyle=\scriptsize\ttfamily,
                emphstyle=\bfseries,
		moredelim=**[is][\color{red}]{@}{@},
		escapeinside= {(*@}{@*)}
	}
\begin{lstlisting}
private void (*@{\color{red}{declareGrouping}@*)(BoltDeclarer boltDeclarer, Node parent, String streamId, GroupingInfo grouping) {
  // the old inconsistent method name is declareStream
  if (grouping == null) {
     boltDeclarer.shuffleGrouping(parent.getComponentId(), streamId);
  } else {
     grouping.declareGrouping(boltDeclarer, parent.get ComponentId(), streamId, grouping.getFields());
  }
}
\end{lstlisting}
\vspace{-0.1in}
\caption{An Example of Inconsistent Method Name}
\label{fig:motiv_1}
\vspace{-7pt}
\end{figure}

Figure~\ref{fig:motiv_1} shows an example in \code{Apache Storm}
project having the method \code{declareGrouping} with an inconsistent
name. It is used to declare a group information for a stream.
%having the inconsistent method name in the method
%\code{declareGrouping}, which is used to declare a group information
%for the stream.
In an earlier version, the method was given the name
\code{declareStream}, which was deemed to be confusing and
inaccurately reflecting the functionality of this method.
%Connecting
%to a stream is only a step in the task of this method.
Therefore, in a later version, a developer performed refactorings to
rename the method into \code{declareGrouping} and at the same time
performed code partition.

This example shows a common case in which during the course of
software development, the name of a method has become confusing and
inconsistent with its functionality.  
%Inconsistent method names could make developers confused, leading to errors due to improper and
%incorrect uses of the method.
Thus, an automated tool to detect inconsistent method names is helpful
for developers to avoid confusing and mistakes.
%during the software development process.

%This example shows a real case during the program developing process
%that the method may have an inconsistent method name. If we want to
%help developers to solve this problem, we need to check the method
%consistency first.

%The old inconsistent method name is \code{declareStream} and this name
%could be confusing for other developers to understand the function of
%this method, because all steps inside of this method are dealing with
%group information even they may connect to the streamID. So in the
%later version, the developers rename this method with the new method
%name \code{declareGrouping} while doing the code standardization and
%code partition combination. This example shows a real case during the
%program developing process that the method may have an inconsistent
%method name. If we want to help developers to solve this problem, we
%need to check the method consistency first. If it is inconsistent, we
%need to generate a consistent method name in order to help solve this
%problem.

\begin{figure}[t]
	\centering
	\renewcommand{\lstlistingname}{Method}
	\lstset{
		numbers=left,
		numberstyle= \tiny,
		keywordstyle= \color{blue!70},
		commentstyle= \color{red!50!green!50!blue!50},
		frame=shadowbox,
		rulesepcolor= \color{red!20!green!20!blue!20} ,
%                xleftmargin=1.5em, xrightmargin=1em, aboveskip=1em,
%		framexleftmargin=0.5em,
%                numbersep=0em,
%		language=Java,
%               basicstyle=\scriptsize\ttfamily,
                xleftmargin=1.5em,xrightmargin=0em, aboveskip=1em,
		framexleftmargin=1.5em,
                numbersep= 5pt,
		language=Java,
                basicstyle=\scriptsize\ttfamily,
                numberstyle=\scriptsize\ttfamily,
                emphstyle=\bfseries,
		moredelim=**[is][\color{red}]{@}{@},
                escapeinside={(*}{*)}
	}
	\begin{lstlisting}
 private Dimension calculateFlowLayout(boolean bDoChilds){
	...
	if (getParent()!=null && getParent()... JViewport) {
	    JViewport viewport = (JViewport) getParent();
	    maxWidth = viewport.getExtentSize().width;
	} else if (getParent() != null){
	    maxWidth = getParent().getWidth();
	} else {
	    maxWidth = getWidth();
	}
	...
	Dimension d = m.getPreferredSize();
	...
}
	
 public Dimension (*\fbox{XXXXXXXXXXXXXXXX}*)() {
	// The consistent method name is getPreferredSize
	return calculateFlowLayout(false);
}
	\end{lstlisting}
        \vspace{-0.1in}
	\caption{An Example of Method Name Suggestion}
	\label{fig:motiv_2}
%	\vspace{-8pt}
\end{figure}

When the method name is identified as inconsistent, it is also useful
to have a tool to recommend a new name for the method.~There are
several factors that a tool can leverage to suggest a new name for the
method. First, a tool can rely on the body (i.e., the content) of the
method to suggest its name. Second, the types and names of the
arguments and return type of the method could also be used to predict
the method's name. The first and second factors are referred to as the
{\bf internal} content and the {\bf interface} of the method under
study.  These two factors represent the only two key sources of
information that the state-of-the-art approaches have been using for
method name checking/suggestion. Liu {\em et al.}~\cite{kim-icse19}
use clone detection on the methods' bodies to search for similar
methods to suggest similar names. Alon {\em et al.}~\cite{alon_popl19}
also rely on the method's content, however, explore code structures by
using embeddings built from the paths over the abstract syntax tree
(AST) of the method under study. Allamanis {\em et
  al.}~\cite{Allamanis_15,Allamanis_16} and Nguyen {\em et
  al.}~\cite{icse20-methodname} also make use of the method's body,
method interface, and class name, however, break down the names of program
entities into sub-tokens, and then use them to
suggest the method name via neural network models~\cite{Allamanis_16,icse20-methodname}
or a clustering algorithm in the vector space~\cite{Allamanis_15}.

%Nguyen {\em et al.}~\cite{icse20-methodname} also use the name of the
%enclosing class of the method as a factor for method name
%checking/suggestion.

%rebuttal ref

%[1] Allamanis, Miltiadis, Hao Peng, and Charles Sutton. "A
%convolutional attention network for extreme summarization of source
%code." In International Conference on Machine Learning,
%pp. 2091-2100. 2016.

%[2] M. Allamanis, E. T. Barr, C. Bird, and C. Sutton, ¡§Learning
%natural coding conventions,¡¨ in Proceedings of the 22nd ACM SIGSOFT
%International Symposium on Foundations of Software Engineering. ACM,
%2014, pp. 281¡V293.

Despite their successes, the state-of-the-art approaches do not work
for the methods with little contents in their bodies. The method at
line 16 in Fig.~\ref{fig:motiv_2} in \code{Tina POS} project is named
\code{getPreferredSize}.
%was originally named \code{UNKNOWN}, which is confusing to
%developers. Thus, in a later version, a developer renamed it into
%\code{getPreferredSize}.
%
The method contains a single call to \code{calculateFlowLayout}.
Assume that one wants to use an existing name suggestion tool
for the body of this method.
However, the existing approaches relying on the method's body or
interface do not work because 1) none of the tokens of the correct name
(\code{getPreferredSize}) appears there; 2) the code structure in the
body does not help in predicting the method's name. Our tool suggests
the correct name \code{getPreferredSize}, while
MNire~\cite{icse20-methodname} uses the body to suggest the
name~\code{getFlow}. {\em Thus, simply using the method's body and
  interface is not sufficient.}

%\vspace{-0.08in}
\subsection{Key Ideas}\label{keys}

\subsubsection{{\bf Show Me Your Friends, I'll Tell You Who
You Are}}

%Instead of relying on the method's body and interfaces,
%{\tool} is based on that principle.
%that ``Show Me Your Friends, I'll Tell You Who You Are''.
%That is,
In addition to the method's body and interface, we characterize a
method by the surrounding methods that interact with the method under
study.  In this problem, the surrounding and neighboring methods of a
method $m$ under study could include the methods that are called
within the body of $m$ (let us call them {\em callees}), the methods
that are calling $m$ ({\em callers}), the methods within the same
class with $m$ ({\em siblings}), and the program elements declared in
the enclosing class of $m$. For method name consistency checking, in
addition to the method's body and interface, we could use all of those
neighboring methods. For name suggestion, we could use callees
and siblings since the callers of $m$ might not be written yet at
the time that the current method $m$ is being edited.

In this example, while the content of $m$ is short, the callee
context, i.e., the body of the method \code{calculateFlowLayout},
contains sufficient information for name suggestion. In
Fig.~\ref{fig:motiv_2}, examining the body of the callee method
\code{calculateFlowLayout}, we can see that the sub-tokens of the
consistent method name \code{getPreferredSize} appear in the names of
the program entities in the callee.  First, the sub-token \code{get}
appears at lines 3-7, 9, and 12. Second, the sub-token
\code{Preferred} appears at line~12. Third, the sub-token \code{Size}
appears at line 5 and line 12. For consistency~checking, callers and
sibling methods can be used since they might be available.
In general, the contexts complement to one another and to the internal
content of the method, contributing to name suggestion. With the
nature of source code, the case of little internal content and
little contexts of a method is rare.

\subsubsection{\bf Representation Learning from Multiple Contexts}
Our model learns the representation to integrate names of
variables, fields, method calls from multiple contexts. In addition to the
method's body and interface (we call it {\em internal context}), we
also consider the {\em interaction context}, which includes all the
methods interacting with the current method~$m$, i.e., caller methods
(if available) and callee methods. In Fig.~\ref{fig:motiv_1}, the two
sub-tokens \code{declare} and \code{Grouping} in the consistent method
name \code{declareGrouping} can be found at line 4 and~line 6. We also
use the sibling methods in the same class ({\em sibling context})
because they provide the tasks with the same~theme.

%in the class.

%{\tool} parses the code and builds the vector representations for the
%contexts and combine them.

%for consistency checking and name suggestion.

%to check the consistency and to derive the suggested name for the
%method.

%With the two observations, we know that the consistent method name
%could relate to method context directly or related to the interacting
%methods. We collect all features from the current method and the
%interacting methods.

%And we develop an algorithm to combine and learn the representation
%for the current method in order to do the consistency checking and
%method name recommendation.  For example, as in figure
%\ref{fig:motiv_2}, we collect the \code{UNKNOWN} method context and
%the context of the interacting method \code{getPreferredSize} to learn
%the interaction representation and use it to do the further work. The
%details about which specific types context we collect and how our
%model work we will explain later.

%\noindent {\bf [Context Breaking Down with Different Weights]}:

%Because different contexts will contribute differently to {\tool},

Different contexts might have the sequences of sub-tokens with
different lengths and nature. For example, in some cases, those
sequences for callers/callees might positively contribute or
negatively impact in deriving the method name. To help our model learn
the importance of different contexts in different situations, we use a
learning scheme for the weights of the contexts.
%we use a weighted approach to automatically learn different weights
%for them.
In Fig.~\ref{fig:motiv_2}, we collect the internal context including
the body at line 18 and the method return type
\code{Dimension} at line 16 as one type of context. We collect the
name of the method \code{calculateFlowLayout}. Also, we collect the
body~and interface of the method \code{calculateFlowLayout} as the
interaction context.

%Because we need to deal with different contexts at the same time, the
%features in our approach are mixed. Different context will have
%different contribution for the consistent checking and method name
%generation. In order to distinguish them, we break down context into
%different types and give different weights for them in our model.  For
%example, in figure \ref{fig:motiv_1}, we collect the method context
%which include method body in line 18 and the method return type in
%line 16 as one type of context. And then we collect the method name of
%method \code{calculateFlowLayout} which has been called by method
%\code{UNKNOWN} as the second type of context. Also, we collect the
%body of method \code{calculateFlowLayout} as the third type of
%context. We give them different weights in our model to do the
%training and testing. The details about how to add weights will be
%introduced later in details.

\subsubsection{\bf Sub-token Copy and Non-copy Mechanisms}

It is reported that the high percentage of sub-tokens of the method
names can be found in the set of the sub-tokens of the~program
entities in the context. Thus, we leverage this by using a~copy
mechanism for Recurrent Neural Network (RNN). This emphasizes on the
likelihood of directly copying a sub-token from the input into the
output position following a sub-token.
%and the possibility of predicting the output without copying.
%For example, in Figure~\ref{fig:motiv_1}, to generate the name for
%method \code{declareGrouping}, the copying mechanism helps find the
%sub-tokens \code{declare} and \code{Grouping} at line 4 and line 6.

We also design a {\em Non-copy} mechanism that complements to the copy
one. As copying at a sub-token position, the non-copy mechanism
estimates the likelihood score that {\em a certain sub-token $s$ must not
be copied to follow the current sub-token $c$}.
%
%A higher score implies that the probability of {\em not seeing a
%certain sub-token following another one} is higher.
A higher score implies that the occurrence probability of the sequence
consisting of $c$ following $s$ is lower.
Thus, it helps our model push lower in the resulting ranked list the
incorrect candidates with those sub-tokens that must not follow
certain ones. The correct candidates thus are pushed higher in the
list, improving top-ranked accuracy (Table~\ref{toplist}).
For example, as deriving the sub-token at the second position
in the method name, {\em Non-copy} pushes the candidates 
\code{declareSchool} or \code{declareStudent} lower in the list since
\code{School} and \code{Student} have never been seen to follow \code{declare}
in the training set. Thus, the correct option \code{declareGrouping}
will be ranked higher.
%will help give the lower probabilities to other sub-tokens except the
%sub-token \code{Grouping}.  That is because when training, all other
%sub-tokens do not follow the sub-token \code{declare} and the
%sub-token \code{Grouping} often does. It can help the copy model find
%the correct sub-token \code{Grouping} for the second position easier.

Importantly, instead of directly using RNN encoder-decoder, we modify
its operations to integrate our newly developed {\em Non-copy}
mechanism with the {\em Copy} mechanism to help better predict the
method name.
The {\em Copy} and {\em Non-copy} mechanisms complement each
other. Assume that the currently predicted name has two sub-tokens:
$A$ $B$. {\em Copy} mechanism suggests $C_1,C_2,...,C_n$ (from the
input) to likely follow $A$ $B$. That is, {\em Copy} mechanism
suggests $C_1$ is likely to be copied, and $C_n$ is less. However,
{\em Non-copy} mechanism can suggest that $C_2$ (from the input) must
not follow $B$ because in training, it has never seen $B$ $C_2$. Thus,
with only {\em Copy} mechanism, $C_2$ is ranked 2nd, however, with
both, we lower the ranking of $C_2$.

%Tien

%We will explain both mechanisms later.

%Some method name may directly use some sub-tokens from other methods
%name or method body as part of the name. In this way, we build a
%Sub-token Copying and Non-copying Network to help generate the correct
%method name.

%This network calculate the possibility of directly copying some
%sub-tokens as part of the output and the possibility of predicting the
%output without copy.

%$The non-copy here is contributing to the copy part. It means that
%when doing the copy at each sub-token position, we will give lower
%possibility of some sub-tokens because they hardly follow the previous
%sub-token during the training. This can help reduce the search space
%of the copy part and improve the accuracy.

%For example, in figure \ref{fig:motiv_1}, if we want to generate the
%method name for method \code{declareGrouping}, the copying and
%non-copying network can help to find the sub-tokens \code{declare} and
%\code{Grouping} in both \textit{line 4 and line 6}. And copy the
%ub-tokens to the method name.

\section{Concepts and Approach Overview}

%Our approach uses the features from different contexts to check the consistency and generate method names.

%Let us define the important concepts used in our approach.

\begin{Definition}[{\bf Caller and Callee Methods}]
  A caller method (caller for short) is a method calling the
  current method under study for consistency checking or name
  suggestion. A callee method (callee for short) is a method called within the body of the current method.
 
%The current method is the method we want to check consistency or
%generate method name; a caller method is the method that calls the
%current method; a callee method is the method that is called by
%current method.
\end{Definition}

\begin{Definition}[{\bf Contexts}]
We consider four types of context:
%In our approach, we consider the following types of context:
	
{\em \bf 1. Internal Context}: The internal context for the
current method contains the content in the {\bf body}, {\bf types} and {\bf names}
of the {\bf arguments} in the {\bf interface}, and the method's {\bf return~type}.

{\em \bf 2. Interaction Context}: The interaction context for the
current method includes 1) the {\bf names} of the {\bf callers} of the
current method, 2) the {\bf contents} of the callers (i.e., {\bf bodies},
{\bf interfaces}, {\bf return types}), 3) the names of the {\bf callees}, and
4) the contents of the callees (bodies, interfaces, return types).

%  The information include 1) method names of caller methods, 2) method
%  content of caller methods, 3) method names of callee methods, and 4)
%  method content of callee methods. Method content here include method
%  return types, interfaces, and method bodies. We consider there are
%  four small kinds of context included.
	
{\em \bf 3. Sibling Context}: The sibling context for the current
method $m$ includes 1) the {\bf names} of the methods in the same class with
$m$, and 2) the {\bf contents} of sibling methods (including the {\bf bodies},
{\bf interfaces}, and {\bf return types}).

%The information include 1) method names and 2) method content of all
%other methods in the same class as the current method. Method content
%here include method return types, interfaces, and method bodies. We
%consider there are two small kinds of context included.

{\em \bf 4. Enclosing Context}: The enclosing context for the current
method includes 1) the name of the enclosing class of the current
method and 2) the names of the program {\bf entities},~{\bf method calls}, {\bf field
accesses}, {\bf variables}, and {\bf constants} in the~class.
        
\end{Definition}

\begin{figure}[t]
	\centering
	\includegraphics[width = 3.6in]{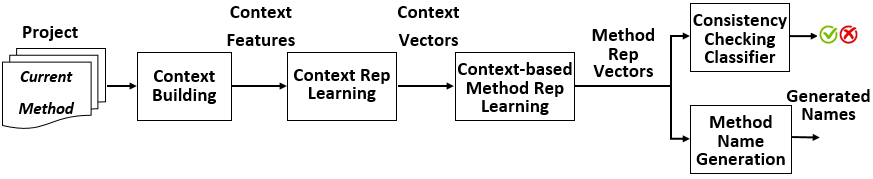}
%	\vspace{-0.12in}
	\caption{Overview of {\tool}'s Architecture}
	\label{Fig:overview}
%	\vspace{-15pt}
\end{figure}
Figure~\ref{Fig:overview} shows {\tool}'s overview.
%for both method name inconsistency checking and suggestion.
%The input is source code of the current method in a project.
%The output has two folds: 1) {\tool} can decide if the current method's name is
%consistent or not, and 2) It can suggest a proper name for the
%current method.
It is aimed to~help in two usage scenarios. First, the project is in
the development process and the source code of the current method
under~consideration and the source code of the callers, callees,
siblings, and the enclosing class are available. {\tool} can be used
to determine the consistency of its name, and then it is used to
suggest an alternative name if the current name is deemed
inconsistent. In the second scenario, the source code of the current
method is written, however, its caller methods are~not available since
its name is not there yet. {\tool} can be used to suggest a proper
name for that method in this scenario.

There are four main steps in {\tool}:

%\paragraph{{\bf 1. Context Building}}

\vspace{0.01in}
{\bf 1. Context Building.}  We collect internal context, interaction
context, sibling context, and enclosing context to build context
features. Specifically, for each type of context, we extract the names
of the program entities appearing in the context. We use the
CamelCase and Hungarian convention to break each name into a sequence
of sub-tokens. For example, \code{calculateFlowLayout} is broken into
\code{calculate}, \code{Flow}, \code{Layout}. We use the sequence of
all the sub-tokens of the program entities' names in the
same order in the source code as the feature for that
  context. The rationale of using the sequence of sub-tokens as
feature (instead of PDG/AST) is the naturalness of
names~\cite{icse20-methodname}:~developers do not give random
names; they use names of program entities or method calls/fields
relevant to the task of the current method.

%from the result of the empirical study by Nguyen {\em et
%  al.}~\cite{icse20-methodname}. They reported that for the problem of
%method name suggestion and consistency checking, the sequence of
%sub-tokens is more important than source code
%structure~\cite{icse20-methodname}.

%First of all, we have a project with the current method $Method_{current}$ to start with. In order to learn the useful information for the current method to do the consistency checking and method name generation, we build different types of contexts for the current method $Method_{current}$. We collect enclosing context, internal context, interaction context, and sibling context to build context features. All context features $F_i$ we built in this step would be the output of this step.

%\paragraph{{\bf 2. Context Representation Learning}}

\vspace{0.02in}
{\bf 2. Context Representation Learning.}
After extracting as features the sequences of sub-tokens for the
contexts, we need to convert those sequences into vector
representations for the models in the later steps. We
use a word embedding model to convert the sequences of sub-tokens
into the vectors.
%Word embedding allows {\tool} to learn the representations of the
%contexts needed for the name suggestion and consistency checking
%components later.
We put the vectors for all the sub-tokens
in the order of the appearances of those sub-tokens in the source
code. As the result, for each context, we have a
sequence of representation vectors corresponding to the sub-token
sequence of that~context.

%For each context feature $F_i$, we need to learn the representation of it before we put them into our model. We use the word embedding tool GloVe \cite{} on all context features of all methods in the training dataset to get the sub-token representation vector $V_{i,j}$ for each sub-token $S-T_j$ in each context feature $F_i$. By putting all sub-token representation vector $V_{i,j}$ together, we will have a set of representation vectors $Vec_i = [V_{i,1}, ..., V_{i,j}]$ for the context feature $F_i$ that is the output of this step.

%\paragraph{{\bf 3. Context-based Method Representation Learning}}

%\vspace{0.03in}
{\bf 3. Context-based Method Representation Learning.}
%To determine the consistency or to suggest a name for the current
%method $m$,
{\tool} relies on all four contexts (internal, interaction, sibling,
enclosing). Thus, we need to produce a representation for method $m$
with the encoding of all contexts. From the previous step, for a
context, we have a sequence of representation vectors. At this step,
we use an encoder to learn the encoding of the features for a
context. We then use a decoder to combine the encoded information for
all the contexts into a representation for $m$, which has the
integration of all contexts.

\vspace{0.03in}
{\bf 4. Consistency Checking and Method Name Suggestion.}
At the last step, {\tool} has two separate models for two tasks.
%consistency checking and method name recommendation.
For consistency checking, it takes as the input the representation of
the method $m$ obtained from the previous step with the existing
method name to be checked, and feeds to a two-channel CNN-based
model~\cite{twochannel} acting as a classifier to classify the
method name to be consistent or not. For method name suggestion, each
vector in the sequence of method representation vectors is a sub-token
representation vector. We use our vocabulary to roll back all the
vectors into the sub-tokens and put them together to generate the
method name.

%The last step of our approach is to do the consistency checking and method name generating based on the method representation vectors $Vec_{current}$ we got in the last step. As for the consistency checking, we use a Convolutional layer \cite{} with SoftMax as classifier to classify the method name to be consistent or inconsistent. As for the method name generating, each vector in the method representation vectors $Vec_{current}$ is a sub-token representation vector. We use our vocabulary dictionary to roll back all vectors into sub-tokens and put them together to generate the method name.

%\section{\tool: Consistency Checking and Name Generation Model}

\section{Context Representation Learning}

\subsection{Context Building}

%First of all, in order to analysis the current method, our approach build the context features first in this step. We regard the whole project as input.

Our first step is to build the contexts from source code.  For the
internal context, sibling context, and enclosing context, we directly
extract the names from the program entities, the return type, and the
types in the interface.  The names are broken into sub-tokens, which
are collected into the sequence in the same appearance order in source
code. The sequences of those sub-tokens represent the contexts. For
example, for the method in Fig.~\ref{fig:motiv_2} (line 16), the
internal context is modeled by the sequence of sub-tokens:
\code{calculate} \code{flow} \code{layout} \code{dimension}. The
internal context also includes the sub-tokens in the return type, the
types and names of the parameters of the method.  The trivial
(sub)tokens with a single character are removed.

%And then we use the Camelcase and Hungarian convention to break down names and types into sub-tokens to have five sequence of sub-tokens for the current method content, the names and contents of sibling methods, and the names and properties of the class as context features.

For the interaction context, we build a call graph using
Soot~\cite{soot}. We then identify the callers and
callees for the current method under study. The sequences of
sub-tokens are built in the same manner as in the previous contexts to
form the feature for the interaction context. For consistency
checking, both callers and callees are considered. However, for method
name suggestion, we consider only the callee methods if
the caller methods do not exist yet.
%because the callers of the current method often do not exist yet.
In Fig.~\ref{fig:motiv_2}, the callee part of the interaction
context includes the sub-tokens built from parsing the method
\code{calculateFlowLayout}:

\code{get} \code{parent} ... \code{view} \code{port} ... \code{max} \code{width} \code{dimension} \code{get} \code{prefer} \code{size} \code{boolean} \code{dimension}

Other callees are processed in the same manner. 
All the sequences of sub-tokens for all contexts are used as input in
the next step. We denote those sequences as {\em context features}.

\subsection{Context Representation Learning}

%When we start to do the consistency checking and method name generation on the current method $Met_{current}$, we have the content features $F_i$ from last step as input to learn the useful information.

%To make the sequences of sub-tokens obtained from last step usable in
%our deep learning model in the later steps, we need to generate the
%representation vectors for the contexts.

To convert the sequences of sub-tokens into vectors, we consider all
the sequences of sub-tokens for all the contexts as the sentences of
words. We use {\em GloVe}~\cite{glove2014}, a word embedding technique, to
produce the vectors for the collected sub-tokens. We use {\em GloVe},
instead of {\em Word2Vec}, due to its capability of learning to
represent the words from the aggregated global word-word co-occurrence
statistics, which captures the relationships between neighboring
sub-tokens.

To build the vector representation for each context, we replace the
sub-tokens in a sequence for the context feature with the
corresponding {\em GloVe} vectors for the sub-tokens. We maintain the same
order as the appearance order in the source code.  For example, for a
context feature sequence $F_i$, we have a sequence of vectors
$V_{F_i}$ = [$v_1$, $v_2$, ..., $v_n$], in which $v_i$ is a {\em GloVe} vector
for a sub-token in the sequence.
Because the sequences of vectors might have different lengths, we
perform zero padding by
%considering the maximum length and
filling the zero vectors for the sequences whose lengths are less than
the maximum length.  This makes the sequences of vectors have the same
length.

\section{Context-Based Method Representation}

%As we mentioned in overview section, our approach use seq2seq
%framework \cite{} to generate interaction representation. In this
%step, let us introduce the model we use first and then talk about the
%how we do the encoding for context features and how we decode context
%features into method representation vectors.

From the previous step, a context is represented as a sequence of
vectors in which each vector represents a sub-token.
%in the names of program entities in the context.
The goal of this step (Fig.~\ref{Fig:approach})
%, the context-based method representation learning,
is to produce a representation for the given method that integrates
all of its contexts.

%Figure~\ref{Fig:approach} shows the model's architecture at this step.

%\subsubsection{\textbf{Encoder and Decoder}}

%\vspace{-0.06in}
\subsection{\textbf{Context Feature Encoding}}

%To achieve that goal, we use encoder and decoder to combine the
%vectors representing the contexts.

%To achieve that goal,
We first use a RNN-based {\em seq2seq} encoder to encode the sets of
vectors for all contexts from the previous step. For each context, we
encode it with a Gate Recurrent Unit (GRU).
We use multiple GRUs because different contexts might have different
structures and types of information.
%
%A single GRU cannot capture all important patterns well at the same
%time.
Multiple GRUs also help reduce the cross influence between different
contexts.

\begin{figure}[t]
\centering
\includegraphics[width = 3.2in]{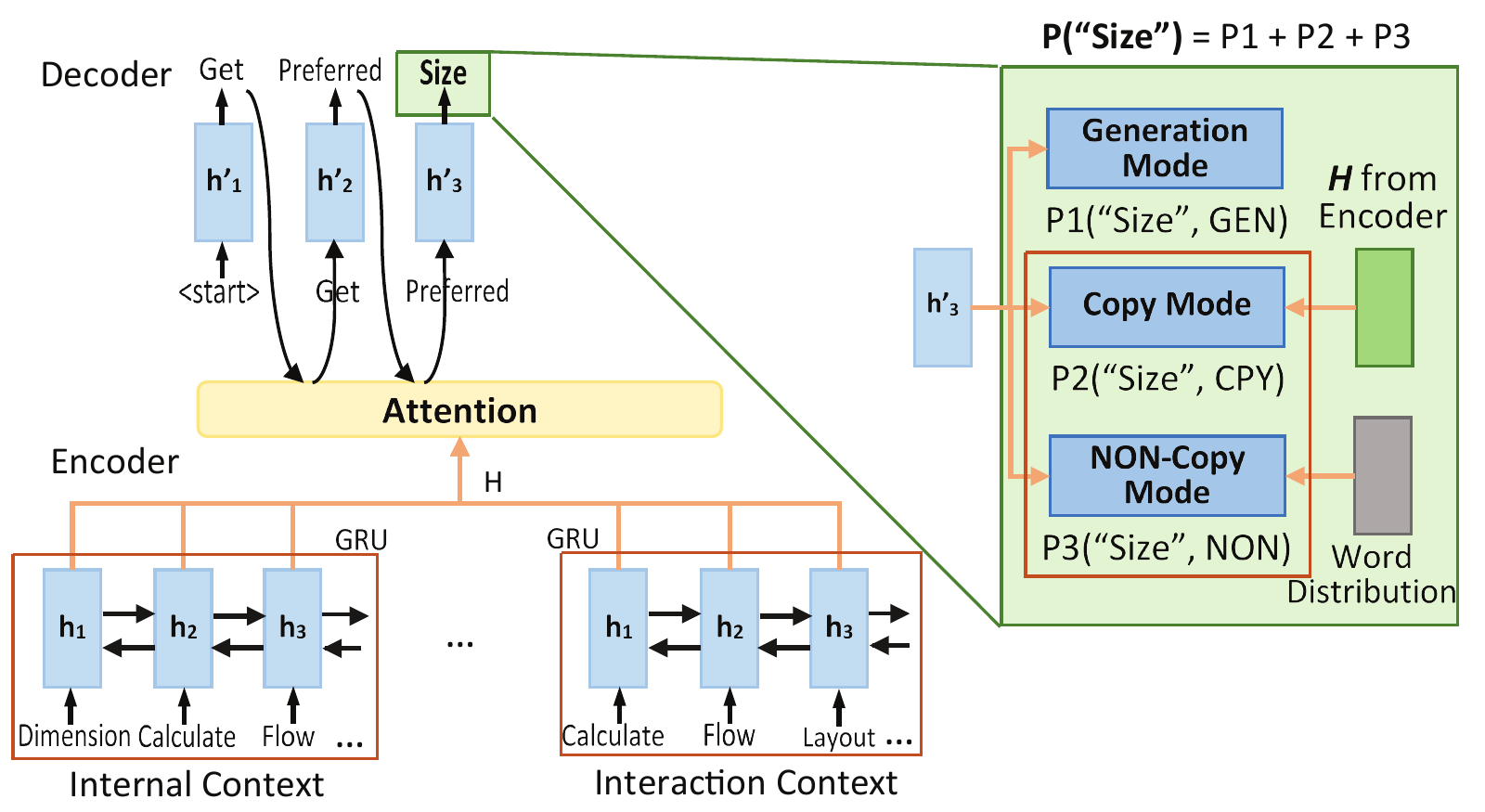}
%\vspace{-0.12in}
\caption{Context-based Method Representation Learning}
\label{Fig:approach}
%\vspace{-0.135in}
\vspace{-5pt}
\end{figure}

%Tien
The input for each GRU$_i$ is the sequence $V_{F_i}$ of $n$ vectors
representing a context. Each vector represents a sub-token in the
names of the program entities in the context. For example, in
Fig.~\ref{Fig:approach}, one GRU is used for the interaction context
including the sub-tokens in the name, body, and interface of the
callee method:
%in our example in Section~\ref{motiv:sec}:
\code{Calculate} \code{Flow} \code{Layout} ....
Another GRU is used for the internal context including the
body/interface of the current
method: \code{Dimension} \code{Calculate} \code{Flow} ...
For each time step $t$, we input one vector $V_t$ in these $n$ vectors
and we get one hidden state vector $h_t$ as the output for this time
step.
By collecting all outputs for each time step, we have a list of hidden
state vectors $H_i = [h_1, ..., h_n]$, which is the output of $GRU_i$.
%

%For example, in figure \ref{fig:approach}, there are serveral different GRUs for different context features, such as one GRU layer to encode current method content $Dimension calculation flow$ and one GRU layer to encode callee method name $calculate flow layout$.
%All hidden states $h_1, h_2, h_3$ arethe  output for the decoder part.

%1) \textbf{RNN-based seq2seq model:} We use an RNN-based seq2seq model
%for such encoding/decoding. In the seq2seq framework,

Mathematically, the input sequence $V_{F_i}=[v_1, v_2,..., v_n]$ is learned
and converted into a hidden vector $H_i$ by the
encoder. The decoder is used to transfer the representation hidden
vector $H_i$ back to the target sequence $Y=[y_1, y_2, ...,
y_m]$. 
\begin{equation}\label{eq:1}
	h_t = f(v_t, h_{t-1})%\:\:\:\:\:\:\:\:\:\:	h'_xt = g(y_{t-1}, h'_{t-1}, H_i)
\end{equation}
\vspace{-10pt}
\begin{equation}\label{eq:2}
	h'_xt = g(y_{t-1}, h'_{t-1}, H_i)
\end{equation}	
\vspace{-10pt}
\begin{equation}\label{eq:3}
	p(y_t|y_1,....,y_t-1,V_{F_i}) = s(y_{t-1}, h_t, H_i)
\end{equation}	
Formula~\ref{eq:1} is for encoder RNN. $h_t$ is the
hidden state in time step $t$; $f$ represents the RNN dynamic
function. Formula~\ref{eq:2} is for the decoder. $h'_t$ is the
hidden state for the decoder at time step $t$; $g$ represents the RNN
dynamic function. Formula~\ref{eq:3} is for prediction: $s$ is
the possibility calculation function.

By putting together all the outputs $H_i$s of all GRUs, we
obtain all the hidden states vectors $H = [H_1, H_2, ...,H_i]$. This
is the output of the encoder and used in the attention layer.
%next in the process.

Because not all the sub-tokens in a context are equally important,
%in deciding the method name,
we aim to put emphasis on certain sub-tokens. Such emphasis is
learned via an {\bf attention mechanism}.
Technically, the attention layer uses a changing context $C_t$
instead of one hidden vector $H_i$. $C_t$ is calculated as
the weighted sum of the encoder's hidden states:
\begin{equation}\label{eq:4}
	C_t = \sum^n_{i=1}{\alpha_{t,i}h_i}  \:\: \:\:\:\:\:	\alpha_{t,i} = \frac{e^{r(h'_{t-1}, h_i)}}{\sum^n_{i'\neq i}{e^{r(h'_{t-1}, h_{i'})}}}
\end{equation}	
%\begin{equation}\label{eq:5}
%	\alpha_{t,i} = \frac{e^{r(h'_{t-1}, h_i)}}{\sum^n_{i'\neq i}{e^{r(h'_{t-1}, h_{i'})}}}
%\end{equation}	
where $r$ is the function used to represent the strength for
attention, approximated by a multi-layer neural network.

%\vspace{-0.08in}
\subsection{\textbf{Context Feature Decoding}}

%We use the GRUs for the decoder

This is our new component (Fig.~\ref{Fig:approach}) in which we modify
the operation of the GRUs at the output layer of the decoder to
integrate our newly developed {\em Non-copy} mechanism. It operates in
connection with the attention layer.

First, at each time step $t$ for a GRU, the previous hidden state
$h'_{t-1}$ is used as the input for the attention layer and the output
of the attention layer will be used as the input of the GRU at the
time step $t$. In Fig.~\ref{Fig:approach}, the vector
for the sub-token \code{Preferred} obtained from the output of the
attention layer at the time step 2 is used as the input of the GRU at
time step 3. This emphasizes on the important sub-tokens
while the decoding is performed at each time step.

Next, at the output layer of the decoder, we also integrate the
operation of two mechanisms: {\bf CopyNet}~\cite{copynet16} and {\bf
  Non-copy}.
%which is a novel component we designed for this work.
%Let us first explain the two mechanisms and how we apply to
%our problem.

%For decoder to do the prediction, each time step $t$ we pick the
%previous hidden state $h'_{t-1}$ and put it to the attention layer and
%use the results come back from attention layer as the input for time
%step $t$. For example, in figure \ref{Fig:approach}, we use the word
%$preferred$ which we get from time step $2$ as input for time step
%$3$.

%After the decoder GRU running, the decoder RNN get the representation vector $V_{RNN}$. Then, we use the generated representation vector $V_{RNN}$, pick the hidden states $H$ and the statistical analysis on word distribution to get the prediction vector $V_t$ based on word possibilities. Which word has higher possibility in the current time step $t$, we pick the word representation vector as the current time step $t$ output. The word prediction possibility is calculated by adding copy-mode possibility , and generation-mode possibility together which are $p("Size", cpy)$ and $p("Size", gen)$ in the green box in the figure \ref{Fig:approach}. The copy-mode possibility $p("Size", cpy)$ is also rely on our non-copy possibility $p("Size", non)$. 

%\vspace{-0.05in}
\subsubsection{{\bf CopyNet}}
\label{copynet}
{\em CopyNet}~\cite{copynet16} is the copy mechanism with the RNN {\em
  seq2seq} model and attention mechanism. It calculates the
possibilities of copying the input sub-tokens to the output.
It has two modes: generation-mode (denoted by \code{GEN}) and
copy-mode (\code{CPY}) for prediction. The attention mechanism with
RNN uses one function. The state update for {\em CopyNet} considers
not only the word embedding, but also the corresponding
location-specific hidden state in the set of RNN encoder's hidden
states $H = [h_1,h_2,...,h_t]$.
\begin{equation}\label{eq:6}
\begin{split}
	p(y_t|h'_t,y_{t-1},C_t, H) = p(y_t, GEN|h'_t,y_{t-1},C_t, H) + \\
        \quad \quad p(y_t, CPY|h'_t,y_{t-1},C_t, H)
        \end{split}
\end{equation}	
\begin{equation}\label{eq:7}
	p(y_t, GEN|.) = \left \{ 
	\begin{array}{ll}
		\frac{1}{Z}e^{u_{GEN}(y_t)} &  y_t \in dic_{all}\\
		0 & y_t \in   \overline{dic_{all}}  \cap dic_{in}\\
		\frac{1}{Z}e^{u_{GEN}(UNK)} &  y_t \notin dic_{all} \cup  dic_{in}\\
	\end{array}
	\right.
\end{equation}
\begin{equation}\label{eq:8}
	p(y_t, CPY|.) = \left \{ 
	\begin{array}{ll}
		\frac{1}{Z}\sum_{j:v_j=y_t}e^{u_{CPY}(v_j)} &  y_t \in dic_{in}\\
		0 & otherwise\\
	\end{array}
	\right.
\end{equation}

$u_{GEN}$ and $u_{CPY}$ are the score functions for the
generation-mode and copy-mode; $Z$ is the normalized term used by two
modes; and $dic_{all}$ and $dic_{in}$ are the overall dictionary and
the dictionary for the input only.

\vspace{0.03in}
\subsubsection{\bf Non-copy mechanism}

%Non-copy mechanism is 
%A novel component that we design in this work.
%based on CopyNet.
%
%We perform the statistic analysis on the word distribution to
We aim to determine the possibility of a sub-token that must {\em not}
follow the current sub-token.
%
%Tien removed
%We add the non-copying mechanism aiming to eliminate the sub-tokens
%with low occurrence possibilities following certain sub-token. That
%is, we reduce the search space of the copying mechanism. Therefore, it
%helps improve the performance.
For instance, if the sub-token \code{get} at the ($m$-$1$)
position is a known one, and the sub-token \code{Preferred} follows
it in the training dataset, we calculate the possibility that
\code{Preferred} will {\em not} follow the sub-token \code{get}.
Such calculation is based on our statistical analysis on the word
distribution on the vocabulary.

Mathematically, for the method name $y_{m-1}$ at the position $(m$-$1)$,
we calculate the possibility that a certain sub-token will not be the
next one at the position $m$. That possibility score, denoted by
$p(y_{m},NON|y_{m-1})$, is computed as
\begin{equation}\label{eq:9}
 p(y_{m}, NON|y_{m-1}) = \left \{
	\begin{array}{ll}
	1 - \frac{count_{(y_m|y_{m-1})}}{count_{y_{m-1}}} & y_m, y_{m-1} \in dic_{all}\\	
	0 & y_{m-1} \notin dic_{all}\\
	1 & others\\
	\end{array}
	\right.
\end{equation}
Where $count_{(y_m|y_{m-1})}$ is the occurrence count for the
sub-token $m$ that follows the sub-token at $(m$-$1)$ in the training
dataset, and $count_{y_{m-1}}$ is the occurrence count for the
sub-token at $(m$-$1)$. This formula is to calculate the word
distribution possibility, which is between [0,1].
%
%We only count the distribution $WD$ on the interaction and sibling
%method names.
%
The larger value means that this sub-token has higher possibility of
{\em not} following the previous sub-token at $(m$-$1)$. Moreover,
because we have multiple context features as input, which can pass to
the copy mode, we add all the possibilities together as the total
copy-mode possibilities with different weights. With this, the
copy-mode Formula~\ref{eq:8} has now become:
\begin{equation}\label{eq:10}
\footnotesize
	p(y_t, NEW|.) = \sum_{i = 1}^{I}W_ip_i(y_t, CPY|.) + W_{NON}p(y_t, NON|y_{t-1})\\
%	 = \left \{
%	\begin{array}{ll}
%		\frac{1}{Z}\sum_{j:v_j=y_t}e^{u_{CPY}(v_j)} + \sum_{i = 1}^{I}W_ip_i(y_{t-1}, NON|y_{t})&  %y_t \in dic_{in}\\
%		0 & otherwise\\
%	\end{array}
%	\right.
\end{equation}
Where $W_i$ is a trainable weight for different types of context features, $W_{NON}$ is a trainable weight and always less than zero, and $I$ is the total number of types of context features.

%To learn the encoding of feature information, we use Gated Recurrent Unit (GRU) \cite{} to learn the encoding for each content feature $F_i$.The reason we use GRU here is that GRU is good at dealing with sequence based information and many existing studies \cite{} proved that GRU can do the encoding well in the seq2seq framework. What's more, in our program, the relationship between the neighbor sub-tokens is important and GRU is good at attracting this kind of relationship. 

%More specifically, GRU each time can process one input vector at time step $t$ based on the previous time step $t-1$ result to encode the input vector. There are the formulas for GRU:

%\begin{equation}\label{eq:1}
%	u_t = sigmoid(W_{u1}x_t + W_{u2}h_{t-1} + b_u)
%\end{equation}	

%\begin{equation}\label{eq:2}
%	r_t = sigmoid(W_{r1}x_t + W_{r2}h_{t-1} + b_r)
%\end{equation}	

%\begin{equation}\label{eq:3}
%	h_t = u_t \odot h_{t-1} + (1-u_t) \odot tanh(W_{h1}x_t + W_{h2}(r_t \odot h_{t-1}) + b_h)
%\end{equation}	

%Where $u_t$ is the update gate vector; $r_t$ is the reset gate vector; $h_t$ is the hidden states/output vector at time step $t$; $W$ are parameter matrices; $b$ are parameter variables; and $\odot$ represent Hadamard product.

%To make different features

%\input{sections/approach_step2_encoding}
\vspace{0.04in}
\subsubsection{\textbf{Method Representation Decoding}}

We have modified the decoder part (see our new component in
Fig.~\ref{Fig:approach})
%of the traditional RNN-based encoder-decoder
to integrate the copy and non-copy mechanisms.
%The box in the right side of Figure~\ref{Fig:approach} illustrates our
%new component.
%
In the traditional GRU for an RNN decoder, the output layer is
computed according to Formula~\ref{eq:2}. We still keep that
computation as one of the three factors to determine the output of the
decoder, which is shown as \code{Generation Mode} in Fig.~\ref{eq:2}.
For example, for the sub-token \code{Size} after \code{Preferred},
that computation is as \code{P(Size,GEN)}. In addition, we also
integrate the {\em CopyNet} mechanism to determine the possibility of
a sub-token based on the sub-token copying as in Formula~\ref{eq:8} in
Section~\ref{copynet}: \code{P(Size, CPY)}. For {\em Non-copy}
mechanism, we integrate with the copy mechanism from
Formula~\ref{eq:9}.  Thus, the new formula for the combination of copy
and non-copy mechanisms is Formula~\ref{eq:10}.

%To compute Formula~\ref{eq:9} for the non-copy mechanism, we rely on
%the computation of word distribution. For the copy mechanism, we use
%the hidden states $H$ from the output of the encoder.

The possibility score of a sub-token as the output of the decoder at a
time step $t$ is the summation of the three factors. Thus, the
possibility score of a sub-token, e.g., \code{w=Size}, is calculated
as \code{P(w)} =
\code{P1(w,GEN)} + \code{P2(w,CPY)} + \code{P3(w,NON)}.
We will pick the representation vector for the sub-token with the
highest possibility score in the current time step $t$ as the output
of the decoder at that time step.

Finally, after having the prediction vector $V_t$ for all time steps,
we put them together to obtain a set of vectors in the original order
as the set of vectors $V_{cur}$ for the current method.

%After the decoder GRU running, the decoder RNN get the representation
%vector $V_{RNN}$. Then, we use the generated representation vector
%$V_{RNN}$, pick the hidden states $H$ and the statistical analysis on
%word distribution to get the prediction vector $V_t$ based on word
%possibilities. Which word has higher possibility in the current time
%step $t$, we pick the word representation vector as the current time
%step $t$ output. The word prediction possibility is calculated by
%adding copy-mode possibility , and generation-mode possibility
%together which are $p("Size", cpy)$ and $p("Size", gen)$ in the green
%box in the figure \ref{Fig:approach}. The copy-mode possibility
%$p("Size", cpy)$ is also rely on our non-copy possibility $p("Size",
%non)$.

%As for the decoder side, we do the similar thing as \textit{CopyNet}. We use one GRU layer as decoder to learn the prediction results. The only different is that we add the non-copying possibility to the copy-mode when do the prediction. 

%For decoder to do the prediction, each time step $t$ we pick the previous hidden state $h'_{t-1}$ and put it to the attention layer and use the results come back from attention layer as the input for time step $t$. For example, in figure \ref{Fig:approach}, we use the word $preferred$ which we get from time step $2$ as input for time step $3$. 

%\vspace{-0.08in}
\section{Consistency Checking and Name Suggestion}

%After having the set of vectors $V_{cur}$ for the
%method $m$, {\tool} performs consistency checking or suggestion.

%\vspace{-0.08in}
\subsection{\textbf{Consistency Checking}}

To check whether the method name is (in)consistent, we use a
two-channel CNN model~\cite{twochannel} as a classifier on the set of
vectors $V_{cur}$, which can be viewed as a matrix $M_{cur}$.
%
%The CNN-based two-channel model has been shown to be able to perform
%well in the image classification problem, we use it on $M_{cur}$.
To build the second channel, we apply the same word embedding step to
represent the name of the current method as the set of vectors
$V_{exist}$. We consider that set as the matrix $M_{exist}$ and
combine with the matrix $M_{cur}$ to form the two-channel 
representation matrix $M_{classification} = [M_{cur}, M_{exist}]$,
which is fed to the two-channel CNN model.
%The new matrix is used as the input for the two-channel CNN model.
The output is produced by the {\em softmax} function. The value is
between [0-1] in which 1 represents consistency and 0 represents
inconsistency.

%As for consistency checking, a classifier is necessary in our approach to classify the consistent and inconsistent method names. In our study, we use a CNN based 2-channel model\cite{} as classifier to do so. We use it here because the 2-channel model can learn the similarity between our method representation and existing method name that is used to check the existing method name consistency. And 2-channel model has been proved to be good on image classification problem which is similar as our problem.

%As for the details, the input of the classier is the method representation vectors $V_{current}$ which we regard these vectors together as a matrix $M_{current}$ by considering each vector as a raw and link them in the original order. And also we use the existing method name as input here too. We use the same learned word embedding in \textit{step2} to represent all sub-tokens in the existing method name. By doing this, we have the existing method name representation vector set $V_{existing}$ which also could be regarded as a matrix $M_{existing}$. By combining $M_{current}$ and $M_{existing}$ in a new axis, they become a 2-channel image representation matrix $M_{classification} = [M_{current}, M_{existing}]$ that is the input for the 2-channel model. The output is produced by SoftMax that the value between 0 and 1 where 1 represent consistent and 0 represent inconsistent.

%\vspace{-0.08in}
\subsection{\textbf{Method Name Suggestion}}

%As for generating the method name, it is easier than the consistency checking.

We consider the sequence $V_{cur}$ produced by the previous step is
the vector representations for the method name under
study. Specifically, we consider each vector $V_k$ as the
representation for a sub-token in the suggested name of the
method. From the dictionary $dic_{all}$ for all the vocabulary in the
corpus, we find the closest vector $V_{s_k}$ to the vector $V_k$ and
use the sub-token $s_k$ corresponding to $V_{s_k}$ as the suggested
sub-token for $V_k$.  Finally, we obtain the sequence of the
sub-tokens for all $V_k$s.  The resulting sequence is considered as
the suggested name for the method under study. The order of the
sub-tokens is the same as the order of the vectors $V_k$s in the
sequence of the representation vectors for the method $V_{cur}$.

%As we already have the overall dictionary $dic_{all}$, for method
%%representation vectors $V_{current}$, each vector we regard it as a
%sub-token representation vector $V_k$ for a sub-token position. By
%checking the overall dictionary $dic_{all}$, we replace back the
%sub-token vector $V_k$ to the real world sub-token $k$. By replacing
%all sub-token representation vectors into real world sub-tokens, our
%approach can link these sub-tokens together with the original order to
%generate method name $Name_{generated}$ for the current method
%$Method_{current}$ that is the output of this step.

%\begin{figure}[t]
	%\centering
	%\includegraphics[width = 3.3in]{graphs/approach_example_2.png}
	%\vspace{-0.08in}
	%\caption{Context Building and Context Representation Learning}
	%\label{Fig:approach_example_2}
%\end{figure}

%\vspace{-0.08in}
\section{Empirical Evaluation}
\label{eval:sec}

\subsection{Research Questions}
%We answer the following questions:

\noindent\textbf{RQ1.} {\em Method Name Consistency Checking Comparative Study.} 
How well does {\tool} perform in comparison with the state-of-the-art method consistency checking approaches?

\noindent\textbf{RQ2.} {\em Method Name Suggestion Comparative Study.}
How does {\tool} perform in comparison with the state-of-the-art
method name suggestion approaches?

\noindent\textbf{RQ3.} {\em Impact Analysis of Different Contexts and Weighting Scheme.}  How do distinct types of contexts
and their different weights affect the overall performance of
{\tool}?

\noindent\textbf{RQ4.} {\em Impact Analysis of Copy and Non-copy Mechanisms.} 
How do copy and non-copy mechanisms affect accuracy?

%\noindent\textbf{RQ5. The Effectiveness of Adding Different Weights Analysis.} 
%How do different weights added on different contexts affect the overall performance of {\tool}?

\noindent\textbf{RQ5.} {\em Method Name Suggestion Accuracy Analysis.} 
How does {\tool} perform on the un-seen method names and the methods in various~sizes?

\noindent\textbf{RQ6.} {\em Live Study.} 
How does {\tool} perform on the currently active real-world projects?

\subsection{Experimental Methodology}

\subsubsection{\textbf{Datasets}}\label{dataset}

%We use two datasets to evaluate our approach and baselines for method name consistency checking and method name generation tasks.
%We evaluate our {\tool} on two tasks: method name consistency checking and method name suggestion. 

\noindent {\bf Corpus for Consistency Checking.}
For comparison, we used the same dataset from Liu {\em et
al.}~\cite{kim-icse19}, which was also used in another baseline
approach MNire~\cite{icse20-methodname}.  The training dataset
(Table~\ref{consistency_data}) from that corpus was collected from the
highly-rated, open-source projects from four communities, namely {\em
Apache}, {\em Spring}, {\em Hibernate}, and {\em Google}. It contains
the latest versions of 430 Java projects with +100 commits. In total,
it has 2,119,573 methods, which were considered as consistent names
because the methods whose names are stable for a long time were
selected.
%=======
For the testing dataset, Liu {\em et al.}~\cite{kim-icse19} mined the
methods whose names were modified by developers for the reasons of
misleading names. Finally,~in those projects, there are 1,402 methods
with inconsistent names. They randomly chose another 1,402 methods
with consistent~names to form a test dataset with 2,804 methods.

\vspace{3pt}
%\noindent 
{\bf Corpus for Method Name Suggestion.}
For comparison, we used the same dataset as in
\code{code2vec}~\cite{alon_popl19} and MNire~\cite{icse20-methodname},
with 10,222 top-ranked Java projects from GitHub. It has
14,458,828 methods and 1,807,913 unique files.
We split the corpus based on the number of projects, instead of
files. The project-based setting reflects better the real-world usage
of {\tool} where it is trained on the set of existing projects and
used to check for a new project. The overloading/overriding methods
and generated method names were removed.
%Thus, in \textit{MNR} corpus, we split into training and testing
%projects such that the ratios of the numbers of training and testing
%files and methods are comparable to the ratios in the file-based
%setting in \textit{code2vec}~\cite{alon_popl19}.
%We randomly shuffled and split all the projects
%We use the same splitting in MNire~\cite{icse20-methodname} for this
%corpus with 9,772 training and 450 testing projects.

%As for method name generation, we use the dataset from MNire \cite{} directly. The follow table \ref{fig:data_2} shows the data summary.

\if\else
\begin{table}[t]
\caption{Corpus for Method Name Suggestion}
\label{nr_data}
%\small
\begin{tabular}{|l|r|r|r|}
\hline
 & \textbf{Testing data} & \textbf{Training data} & \textbf{Total} \\ \hline
%\multicolumn{4}{|c|}{\textbf{[{\em Dataset 1}] Comparison Experiment with \textit{code2vec}}} \\ \hline
%\#Files & 61,641 & 1,746,272 & 1,807,913 \\ \hline
%\#Methods & 458,800 & 14,000,028 &  14,458,828\\ \hline
%\multicolumn{4}{|c|}{\textbf{[{\em Dataset 2}] Experiments for RQ4, RQ5, RQ6, RQ7}} \\ \hline
\#Project & 450 & 9,772 & 10,222 \\ \hline
\#File & 51,631 & 1,756,282 & 1,807,913 \\ \hline
\#Methods & 466,800 & 13,992,028 & 14,458,828 \\ \hline
%
%
%
%\multicolumn{4}{|c|}{\textbf{[{\em Dataset 3}] Live Study on Real Developers}} \\ \hline
%\#Project  	& 100 &  & 100 \\ \hline
%\#File 	  	& 18,970 &  & 18,970 \\ \hline
%\#Methods 	& 139,827 &  & 139,827 \\ \hline
\end{tabular}
\end{table}
\fi

\iffalse
\begin{table}[t]
	\caption{Corpus for Method Name Suggestion}
	\vspace{-10pt}
	\label{nr_data}
	\small
	\begin{tabular}{|l|r|r|r|}
		\hline
		& \textbf{Test data} & \textbf{Train data} & \textbf{Total} \\ \hline

		\#Project & 1022 & 9,200 & 10,222 \\ \hline
		\#File & 106,247-128,048 & 1,679,865-1,701,666 & 1,807,913 \\ \hline
		\#Methods & 993,715-1,112,302 & 13,346,526-13,465,113 & 14,458,828 \\ \hline

	\end{tabular}
\end{table}
\fi

\vspace{0.03in}
\subsubsection{\textbf{Empirical Procedure}} Let us present our procedure.
%\vspace{5pt}

\noindent\textbf{RQ1.} {\bf Method Name Consistency Checking.}  We
chose the following baselines: 1) {\em Liu {\em et
    al.}}~\cite{kim-icse19}, an IR approach to search for similar
methods to suggest similar names, 2) {\em
  MNire}~\cite{icse20-methodname}, an ML approach using \code{seq2seq}
encoder-decoder on the sub-token sequences in the method's body and
interface.
%We re-ran both baselines.
%
%and cannot reuse the reported numbers since we use different
%machines/environments.
%
We trained each model under study with the same training dataset and
tested it with the same testing dataset.

For hyper-parameter tuning for a model, we used AutoML in
NNI~\cite{NNI} to automatically tune the parameters. We selected the
parameter set that helps a model obtain the highest \code{F-score} and
\code{accuracy}.
For name consistency checking, 2,804 methods are for testing;
90\% (1,907,716 methods) of the training data are for training; and
the remaining 10\% are for tuning.

%10\% of training data (211,957) are for tuning, and the remaining 90\%
%are for training (1,907,716). For name recommendation, 80\% of the
%dataset are for training, 10\% for tuning, and 10\% for testing.

%We ran a model 10 times and get average values.

%\textbf{1) Baselines:} We compared our approach on method name consistency checking task with two baselines: Liu et al. \cite{} and MNire \cite{}

%\textbf{Liu et al.} is the approach that leverages deep feature representation techniques adapted to the nature of each artifact to do the consistency checking.

%\textbf{MNire} is a machine learning approach to check the consistency between the name of a given method and its implementation.

%\textbf{2) Experiment Process:}

%We do the experiment on the Liu et al. dataset for this RQ. We split the data by 90\% and 10\% that 90\% is used to do the training and 10\% data is used to do the testing. We apply all approaches on the same data split for each round training and testing in order to make our research questions can get the fair results for all approaches. As for two baselines, because all of them do not require negative data to do the training. After data split, we only use the method with consistent method name in the training dataset to train their model. But when doing the testing, all three approaches include ours will be tested on all method names in the testing dataset.

\vspace{0.03in}

\noindent\textbf{RQ2.} {\bf Method Name Suggestion.}  We compared with
the following baseline approaches:
%Tien removed Liu's
%1) {\bf Liu {\em et al.}}~\cite{kim-icse19}, 2)
1) {\bf MNire}~\cite{icse20-methodname}, 2) {\bf \code{code2vec}}~\cite{alon_popl19}
3) {\bf \code{code2seq}}~\cite{alon2018code2seq}, and 4) {\bf path-based
  representation}~\cite{Alon_pld18}.
%Tien
%The last three approaches are the code representation models that can
%be used for method name suggestion.
%a \code{seq2seq} encoder-decoder model, 2)
%code2vec~\cite{alon_popl19}, a tool for representing snippets of code
%as continuous distributed vectors that can be used to predict method
%names. We could not compare with Liu {\em et al.}~\cite{kim-icse19}
%for name suggestion due to two reasons.  First, that tool did not work
%on our dataset. We contacted the authors, however, with no response.
%Second, we cannot use the results from their paper for name suggestion
%since their settings (metric, procedure) are different from {\tool},
%code2vec, and MNire.
For comparison, using the same procedure as
MNire~\cite{icse20-methodname}, we randomly split the data by 80\% for
training, 10\% for parameter tuning, and 10\% for testing.

\vspace{0.03in}
\noindent\textbf{RQ3-RQ4.} {\bf Impact Analysis on various
  Components.} We varied our model (e.g., adding each context), and
measured accuracy. We used the parameter settings as in RQ1-RQ2.

%As for RQ3 to RQ5, we are analyzing how different components in our
%model influence its performance. Each time, we pick out one component
%and test the performance of model to get the analysis of that specific
%component.
%As for RQ3, we added each context and study the results.

%For each running, we directly pick the best performed parameter sets
%from RQ1 and RQ2 and keep them the same in order to make the analysis
%clear.

\vspace{0.03in}
\noindent\textbf{RQ5.} {\bf Accuracy Analysis on Method Name
  Suggestion.} We measure the accuracy on un-seen method names and
on the methods with different lengths in the same setting as RQ2.

\vspace{0.03in}
\noindent\textbf{RQ6.} {\bf Live Study.} We use our tool to check the
method names in the active GitHub projects, make pull requests to
rename inconsistent ones, and evaluate the acceptance~responses.

%As for RQ6, we do the evaluation on method name generation
%accuracy. We evaluate the performance of \tool on un-seen method names
%to prove that \tool is generating the method name instead of directly
%store and retrieving the existing names.

%Also, we evaluate our approach on different length of method in order
%to see how our approach extract method name tokens from different
%context. Also, we could compare our approach with MNire in detail to
%see how our approach can be improved by considering new types context
%and using new technologies.

%\noindent\textbf{RQ7. The live study}

%We randomly selected XX active java project from GitHub and run our approach to check the method name consistency. If we found the inconsistent method name, we run our approach to generate a new method name. We randomly get XX generated method name for XX projects and then we created pulling requests for each generated method name.

%We collect the pulling request results as \textit{Accept} which means they think the new name is better and fixed it, \textit{Agree} which means they think the new name is good, but they may cannot change it directly or some other reasons, \textit{Disagree} which means they think the new name is not good, \textit{No-reply} which means there is not a developer to reply our pull request.

\begin{table}[t]
  \centering
\caption{Corpus for Method Name Consistency Checking}
%\vspace{-10pt}
\label{consistency_data}
\small
\begin{tabular}{|l|r|r|}
\hline
 & \textbf{Testing data} & \textbf{Training data} \\ \hline
\#Methods & 2,804 & 2,119,573 \\ \hline
\#Files & -- & 251,362 \\ \hline
\#Projects & -- & 430 \\ \hline
\#Unique method names & -- & 540,547 \\ \hline
%\vspace{-10pt}
\end{tabular}
\end{table}

\begin{table}[t]
\centering
\caption{Corpus for Method Name Suggestion}
\label{nr_data}
\small
\begin{tabular}{|l|r|r|r|}
\hline
 & \textbf{Testing data} & \textbf{Training data} & \textbf{Total} \\ \hline
\#Project & 1,022 & 9,200 & 10,222 \\ \hline
\#File & 51,631 & 1,756,282 & 1,807,913 \\ \hline
\#Methods & 466,800 & 13,992,028 & 14,458,828 \\ \hline
\end{tabular}
\end{table}

\vspace{0.03in}
\subsubsection{\textbf{Evaluation Metrics}}

%We use following evaluation matrix to evaluate our approach:

%\textbf{Consistency Checking: } We use precision, recall, f-score and accuracy to evaluate the model performance. Here are the formulas:

%$$Precision_{IC} = \frac{TP}{TP+FP}$$
%$$Recall_{IC} = \frac{TP}{TP+FN}$$
%$$Precision_{C} = \frac{TN}{TN+FN}$$
%$$Recall_{C} = \frac{TN}{TN+FP}$$
%$$F-score = \frac{2*Precision*Recall}{Precision+Recall}$$
%$$Accuracy =  \frac{TP+TN}{TP+TN+FP+FN}$$

%\textbf{Method Name Generation: } We use precision, recall, f-score and exact match to evaluate the model performance. We regard the true method name as $M$ and the model predicted method name as $N$, then we have the following formulas:

%$$Precision = \frac{subtoken(M)\cap subtoken(N)}{subtoken(N)}$$
%$$Recall = \frac{subtoken(M)\cap subtoken(N)}{subtoken(M)}$$
%$$F-score = \frac{2*Precision*Recall}{Precision+Recall}$$

%The exact match here means that the predicted method name $N$ is exactly the same as true method name $M$.

%\subsection{Procedure and Metrics}

%In general, for each application, we trained {\tool} with the train data. After that, we used {\tool} to predict the name given its contexts.
%Procedure

%%Tien \noindent {\bf Procedure.} For \textit{MCC} application, for a
%%testing method, the predicted class is decided by {\tool} using a
%%varied threshold $T$.
%%We measure the impact of $T$ in our sensitivity study.
For method name consistency checking, we compared the predicted
results against the ground truth on consistent and inconsistent method
names provided as part of the name consistency checking
corpus~\cite{kim-icse19} (Table~\ref{consistency_data}). For name
suggestion, we compared the predicted names by a model against the
good method names in the name suggestion corpus, which was part of
\code{code2vec}~\cite{alon_popl19} (Table~\ref{nr_data}).
%
%We varied $T$ for maximizing $F-score$ of each classification
%(inconsistency or consistency) and Accuracy.
%
%For method name recommending, for each testing method, we compare the recommended name against the original name of the method which is considered as the expected method name given the method's contexts.

For consistency checking, we used the same
evaluation metrics as in Liu {\em et al.}~\cite{kim-icse19} and
MNire~\cite{icse20-methodname} including \code{Precision},
\code{Recall}, and \code{F-score}, for both inconsistency ($IC$) and
consistency ($C$) classes. For $IC$ class, \code{Precision} = 
$\frac{|TP|}{|TP|+|FP|}$, and \code{Recall} = $\frac{|TP|}{|TP|+|FN|}$. For
$C$ class, \code{Precision} = $\frac{|TN|}{|TN|+|FN|}$, and \code{Recall} =
$\frac{|TN|}{|TN|+|FP|}$, in which $TP$ is true positive ($IC$ is
classified as $IC$), $FN$ is false negative ($IC$ is classified as
$C$), $TN$ is true negative ($C$ is classified as $C$), and $FP$ is
false positive ($C$ is classified as $IC$).
For both $IC$ and $C$, \code{F-score} is defined as $\frac{2\times
  Precision \times Recall}{Precision + Recall}$.
For the overall in both $IC$ and $C$, \code{Accuracy} is
defined as $\frac{|TP|+|TN|}{|TP|+|FP|+|TN|+|FN|} = |TP|+|TN|$.

%To measure the accuracy of {\tool} in method name (in)consistency
%checking, we compute Precision, Recall, and F-score for both
%inconsistency ($IC$) and consistency ($C$) classification and Accuracy
%used the same formulas in~\cite{kim-icse19}.
%
%That is, there are four possible outcomes: $IC$ classified as $IC$ (i.e, true positive=$TP$), $IC$ classified as $C$ (i.e., false negative=$FN$), $C$ classified as $C$ (i.e., true negative=$TN$), and $C$ classified as $IC$ (i.e., false positive=$FP$).
%

%\noindent For $IC$,
%$$
%Precision = \frac{|TP|}{|TP|+|FP|}, 
%Recall = \frac{|TP|}{|TP|+|FN|}
%$$
%
%For $C$,
%$$
%Precision = \frac{|TN|}{|TN|+|FN|},
%Recall = \frac{|TN|}{|TN|+|FP|}
%$$

For method name suggestion, we used the same metrics as in
\code{code2vec}~\cite{alon_popl19} and MNire~\cite{icse20-methodname}:
\code{Precision}, \code{Recall}, and \code{F-score} over
case-insensitive sub-tokens. That is, for the pair of an expected
method name $e$ and its recommended name $r$, precision and recall are
computed as: \code{Precision(e, r)}= $\frac{|subtoken(r) \cap
  subtoken(e)|}{|subtoken(r)|}$, and \code{Recall(e, r)}=
$\frac{|subtoken(r) \cap subtoken(e)|}{|subtoken(e)|}$; $subtoken(n)$
returns the sub-tokens in the name $n$.
%
%For example, for a method named \texttt{countLines}, a recommendation
%\texttt{linesCount} is considered as an exact match, a prediction of
%\texttt{count} has a full precision but low recall, and a prediction
%of \texttt{countBlankLines} has a full recall but low precision.
%
\code{Precision}, \code{Recall}, and \code{F-score} of {\em the set}
of the suggested names are defined as the~average values in all cases.
In all experiments, we also counted the exact-matched names
(\code{ExMatch}) and the case-sensitive names.

%%Tien
%%For \textit{MNR}, in all experiments, we also measured the
%%percentages of exact-matched and case-sensitive~ones.

%and ignoring differences in non-alphabetical characters.

%For example, this metric considers \texttt{totalCount} as an exact
%match to \texttt{total\_count}.

\subsection{Experimental Results}

\subsubsection{{\bf RQ1. Method Name Consistency Checking}}

\begin{table}[t]
	\caption{RQ1. Method Consistency Checking Comparison (C: Consistent Methods; IC: Inconsistent Methods)}
	\vspace{-5pt}
        \tabcolsep 4pt
        \small
	\begin{center}
		\renewcommand{\arraystretch}{1} 
		\begin{tabular}{p{0.5cm}<{\centering}|p{1.2cm}<{\centering}|p{2cm}<{\centering}|p{1.4cm}<{\centering}|p{1.4cm}<{\centering}}
			\hline
			 &	& Liu {\em et al.}~\cite{kim-icse19} & MNire~\cite{icse20-methodname} &	{\tool}\\
			\hline
			\multirow{3}{*}{C} & Precision & 46.5\% & 54.1\% & 64.8\% \\
			 & Recall & 68.3\% & 80.6\% & 86.4\% \\  
			 & F-score & 55.3\% & 64.7\% & 74.1\% \\  
			\hline
			\multirow{3}{*}{IC} & Precision & 53.6\% & 60.5\% & 72.3\% \\
			& Recall & 81.3\% & 90.2\% & 92.1\% \\  
			& F-score & 64.6\% & 72.4\% & 81.0\% \\
			\hline
			\multicolumn{2}{c|}{Accuracy} & 56.8\% & 65.7\% & 75.8\% \\
			\hline
		\end{tabular}
		\label{RQ1}
	\end{center}
	\vspace{-10pt}
\end{table}

%As seen in Table~\ref{RQ1}, {\tool} can have higher F-score and accuracy on both consistent methods and inconsistent methods.

As seen in Table~\ref{RQ1}, for {\bf inconsistent method name
detection (IC)}, {\tool} has a relative improvement of (35.0\%, 13.3\%,
25.5\%) and (19.6\%, 2.1\%, 11.9\%) on
(\code{Precision}, \code{Recall}, and
\code{F-score}) in comparison with Liu {\em et al.}~\cite{kim-icse19} and
MNire~\cite{icse20-methodname}, respectively.  We found that for Liu
{\em et al.}~\cite{kim-icse19}, in many cases, the inconsistent
methods might not be classified as inconsistent (lower \code{Recall}), and
the predicted inconsistent methods might be incorrect (lower
\code{Precision}). The main reason is that the principle of ``methods with
similar bodies have similar names and vice versa'' does not hold in
many cases.  For MNire, several inconsistent methods are not
classified as inconsistent (lower recall) since the bodies use similar
sub-tokens, but the methods do not have the same tasks.  For those
cases, {\tool} uses the caller and callee methods to complement for
the internal context in the characterization of a method, and is able
to detect the inconsistencies since it can detect methods with the
same usages but with different names.

For {\bf consistent method name detection (C)}, {\tool} has a relative
improvement of (39.5\%, 26.5\%, 33.9\%) and (19.8\%, 7.1\%, 14.4\%) on
\code{Precision}, \code{Recall}, and \code{F-score} in comparison with Liu {\em et
al.}~\cite{kim-icse19} and MNire~\cite{icse20-methodname},
respectively.

Regarding {\bf accuracy} as considering {\bf both consistent and
inconsistent name detection}, {\tool} relatively improves {\em 33.6\%}
and {\em 15.4\% compared to Liu {\em et al.}~\cite{kim-icse19} and
MNire~\cite{icse20-methodname}}, respectively. We found several
consistent methods with similar bodies, however with different
names. For example, the methods on \code{InputStream}
and \code{OutputStream}, but have the same body of \code{return
stream;}. Both Liu {\em et al.}~\cite{kim-icse19} and
MNire~\cite{icse20-methodname} relies on the body, thus, cannot work
in those cases, while {\tool} distinguishes them via callers/callees
and siblings.  Moreover, there are methods with the same bodies but
names are different due to different enclosing classes intended for
different purposes,
e.g., \code{process.start()}, \code{process.stop()}. The baselines
detected them as inconsistent. {\tool} uses the callers/callees to
recognize its usage, thus, correctly detecting it as~consistent.

%For consistent methods, {\tool} have a relevant improvement by 39.5\%, 26.5\%, 33.9\% and 19.8\%, 7.1\%, 14.4\% on precision, recall, and f-score in comparison with Liu et al. and MNire. For inconsistent methods, {\tool} also have a relevant improvement by 35.0\%, 13.3\%, 25.5\% and 19.6\%, 2.1\%, 11.9\% on precision, recall, and f-score in comparison with Liu et al. and MNire. Also, the {\tool} has a relevant improvement by 33.6\% and 15.4\% on overall accuracy in comparison with Liu et al. and MNire.

\vspace{0.03in}
\subsubsection{{\bf RQ2. Method Name Suggestion}}

%\begin{table}[t]
%	\caption{RQ2. Method Name Suggestion Comparison}
%        \scriptsize
%        \tabcolsep 2pt
%	\begin{center}
%		\renewcommand{\arraystretch}{1} 
%		\begin{tabular}{p{1cm}<{\centering}|p{1.2cm}<{\centering}|p{1.2cm}<{\centering}|p{1.1cm}<{\centering}|p{1cm}<{\centering}|p{1.2cm}<{\centering}|p{1.2cm}<{\centering}}
%			\hline
%			&	code2vec~\cite{alon_popl19} & code2seq~\cite{alon2018code2seq} & Path-based Rep~\cite{alon2018general} & Liu {\em et al.}~\cite{kim-icse19} & MNire~\cite{icse20-methodname} &	{\tool}\\
%			\hline
%                      	ExMatch & 21.7\% & 32.4\% & 23.3\% & & 38.9\% & 44.3\% \\
%			Precision & 60.2\% & 72.3\% & 56.4\% & & 67.4\% & 73.6\% \\
%			Recall & 52.4\% & 66.1\% & 49.2\% & & 63.1\% & 71.9\% \\
%			F-score & 56.0\% & 69.1\% & 52.6\% & & 65.2\% & 72.7\% \\
%			\hline
%		\end{tabular}
%		\label{RQ2}
%	\end{center}
%
%\end{table}

\begin{table}[t]
	\caption{RQ2. Method Name Suggestion Comparison}
	\vspace{-5pt}
        \footnotesize
        \tabcolsep 2pt
	\begin{center}
		\renewcommand{\arraystretch}{1} 
		\begin{tabular}{p{1cm}<{\centering}|p{1.3cm}<{\centering}|p{1.5cm}<{\centering}|p{1.3cm}<{\centering}|p{1.4cm}<{\centering}|p{1.3cm}<{\centering}}
			\hline
			&	code2vec~\cite{alon_popl19}  & Path-Rep~\cite{Alon_pld18} &
			 code2seq~\cite{alon2018code2seq}& MNire~\cite{icse20-methodname} &	{\tool}\\
			\hline
                       	ExMatch & 21.7\%  & 23.3\% & 32.4\% & 38.9\% & {\bf 44.3}\% \\\hline
			Precision & 60.2\%  & 56.4\% & 72.3\% & 67.4\% &  73.6\% \\
			Recall & 52.4\%  & 49.2\%& 66.1\% &  63.1\% &  71.9\%\\
			F-score & 56.0\%  & 52.6\% & 69.1\% & 65.2\% &  72.7\% \\
			\hline
		\end{tabular}
		\label{RQ2}
	\end{center}
	\vspace{-10pt}
\end{table}

As seen in Table~\ref{RQ2}, {\bf 44.3\%} of the cases suggested by
{\tool} at top-1 positions are exactly matched with the
correct method names given by developers. It has relative
improvements from {\bf 13.9\%--104.1\%} compared with the 
baselines.
It also achieves higher \code{F-score} than all the
baselines. Specifically, it has relative improvements of
1.8\%--30.5\% in \code{Precision}, 8.8\%--46.1\% in \code{Recall},
and 5.2\%--38.2\% in \code{F-score} over the baselines.

With regard to \code{ExMatch}, \code{code2vec} and \code{path-based
Rep} have lower values than {\tool} as they mainly encode path-based
contexts with tokens, which have shown as less repetitive than the
sub-tokens~\cite{icse20-methodname}.
\code{code2seq} encodes the path-based contexts as well as the sub-tokens. However, it failed to capture the order of sub-tokens for the exact name recommendation. \code{code2seq} often recommends the relevant sub-tokens, but not in the right order. {\tool} improves \code{code2seq} by 37\%. Also MNire does not consider the callers/callees and siblings, thus it cannot identify more sub-tokens than {\tool}. 

With regard to \code{Recall}, we found
that \code{code2vec} and \code{path-based Rep}
%and Liu {\em et al.}~\cite{kim-icse19}
have lower recall than {\tool} because
the baselines mainly encode path-based contexts within one method. %and thus require two methods to have similar structures to have similar names. In fact, source code with different structures can have similar names. 
\code{code2seq} requires two methods with similar sub-tokens and/or path contexts.
MNire requires two methods with similar sequences of sub-tokens to
have similar names.  {\tool} uses the bodies/interfaces as well as the
interaction, sibling, and enclosing class contexts, thus, is more
flexible.

With regard to \code{Precision}, two methods can be realized in the
same structure, but are named differently since they are in different
classes and are used differently. Because two methods have the
same/similar AST path contexts, \code{code2vec} and \code{Path-based
Rep} suggest the same name, thus, they have lower precisions. MNire
uses the enclosing class but does not consider the interaction and
sibling contexts. Thus, in several such cases, MNire suggests the same
name, while {\tool} suggests the correct name due to the
callers/callees and siblings.

\vspace{0.03in}
\subsubsection{{\bf RQ3. Impact Analysis of Different Contexts and Weights}}
\noindent {\bf A. Context Analysis.}
The base model in this experiment uses only internal context (the
method body and interface). As seen in Table~\ref{RQ3:1}, when adding
the enclosing class of the method
(\code{A+enclosing}), \code{accuracy} increases by 1.1\% (1.7\%
relatively), as both \code{F-scores} for C and IC classes
increase. Considering the sibling methods, \code{accuracy}
additionally increases 2.0\% (3.1\% relatively) as comparing the
columns (B) and (C). Finally, with all the contexts, \code{accuracy}
additionally increases 8.5\%, i.e., 12.6\% relatively. Thus, all
contexts contribute positively toward the overall accuracy.

\begin{table}[t]
	\caption{RQ3. Context Analysis on Consistency Checking}
	\vspace{-7pt}
	\tabcolsep 3pt
        \small
	\begin{center}
		\renewcommand{\arraystretch}{1} 
		\begin{tabular}{p{0.3cm}<{\centering}|p{1.2cm}<{\centering}|p{1.1cm}<{\centering}|p{1.6cm}<{\centering}|p{1.5cm}<{\centering}|p{1.6cm}<{\centering}}
			\hline
			&	& Internal (A) & A+Enclosing (B) & B+Siblings (C) & C+Interaction (\tool)\\
			\hline
			\multirow{3}{*}{C} & Precision & 53.1\% & 54.4\% & 56.5\% & 64.8\% \\
			& Recall & 79.3\% & 80.9\% & 81.9\% & 86.4\%\\  
			& F-score & 63.6\% & 65.1\% & 66.9\% & 74.1\% \\  
			\hline
			\multirow{3}{*}{IC} & Precision & 59.2\% & 61.1\% & 63.4\% & 72.3\%\\
			& Recall & 88.4\% & 89.1\% & 89.3\% & 92.1\% \\  
			& F-score & 70.9\% & 72.5\% & 74.2\% & 81.0\% \\
			\hline
			\multicolumn{2}{c|}{Accuracy} & 64.2\% & 65.3\% & 67.3\% & 75.8\% \\
			\hline
		\end{tabular}
		\label{RQ3:1}
		%		\\
		%		C: Consistent Methods; IC: Inconsistent Methods; INC: Internal Context; ITC: Interaction Context; SC: Sibling Context; EC: Enclosing Context
	\end{center}
%	\vspace{-10pt}
\end{table}

\begin{table}[t]
	\caption{RQ3. Context Analysis on Name Suggestion}
	\vspace{-8pt}
        \tabcolsep 3pt
        \small
	\begin{center}
		\renewcommand{\arraystretch}{1} 
		\begin{tabular}{p{1.5cm}<{\centering}|p{1cm}<{\centering}|p{1.6cm}<{\centering}|p{1.5cm}<{\centering}|p{1.5cm}<{\centering}}
			\hline
			& Internal (A) & A+Enclosing (B) & B+Siblings (C) & C+Interaction (\tool)\\
			\hline
                        ExMatch & 38.3\% & 38.8\% & 39.7\% & 44.3\%  \\
			Precision & 65.7\% & 66.3\% & 68.5\% & 73.6\%  \\
			Recall & 62.4\% & 63.9\% & 65.7\% & 71.9\%  \\
			F-score & 64.0\% & 65.1\% & 67.1\% & 72.7\%  \\
			\hline
		\end{tabular}
%	\\
%		INC: Internal Context; ITC: Interaction Context; SC: Sibling Context; EC: Enclosing Context
		\label{RQ3:2}
	\end{center}
	\vspace{-10pt}
\end{table}

With further analysis, we have observed the following. First, the {\em
enclosing class} provides the context related to the general theme of
the current method, e.g.,
\code{InputStream}~versus \code{OutputStream}. While the method bodies are
the same (\code{return stream;}), {\tool} is able to derive the
correct names \code{getInputStream} and \code{getOutputStream}
by leveraging the enclosing context. Second, the {\em sibling context}
provides the names of the relevant methods to the current one. For
example, in a class that provides mouse handling for a canvas, the
sibling methods \code{onMouseUp} and \code{onMouseDown} give useful
sub-tokens to suggest the method name \code{onMouseOver}.
%The body of the method is more toward the behavioral actions in
%response to the mouse event, and the variables' names in the enclosing
%class are not relevant to the mouse-over action.
Finally, the interaction context helps suggest the names for the
methods with little content in the bodies, e.g., in delegation
methods.
%{\tool}, with all contexts, achieves highest accuracy.
Unlike the existing approaches with only internal context, we leverage
the interactions, siblings, and enclosing contexts of the method as
well as internal context (body/interface) to achieve highest accuracy.

We also performed another experiment to leave one context out and
compare the accuracy with {\tool}'s accuracy in order to determine the
impact of each context. Without the interaction
context, \code{accuracy} and two \code{F-scores} reduce by 11.2\%,
8.5\% and 9.7\%, respectively.  Without the sibling context, the
performance decreases by 8.8\%, 6.5\%, and 7.9\% on accuracy and two
\code{F-scores}, respectively.  The enclosing context also positively
contributes to high performance. In brief, the internal and
interaction contexts contribute the most.

The contributions of contexts are also confirmed by the method name
suggestion results (Table~\ref{RQ3:2}). When adding the enclosing
context to the internal one, \code{F-score} increases by 1.1\%.  When
further adding the sibling context, \code{F-score} additionally
increases by 2.0\%. Finally, adding the interaction
context, \code{F-score} additionally contributes 5.6\%.

%As seen in Table~\ref{RQ3:1}, for method name consistency checking, {\bf all types of context contribute to the overall performance of \tool.} Without either type of context, {\tool} performs worse in every metric.

%Specifically, the Internal Context {\bf (INC)} has the largest impact on \tool. Without it, \tool reduced 33.8\%, 39.7\% and 39.1\% on two f-score and accuracy.

\if\else
\begin{table}[t]
	\caption{RQ3. Context Analysis on Method Name Suggestion}
	\vspace{-11pt}
	\tabcolsep 3pt
	\small
	\begin{center}
		\renewcommand{\arraystretch}{1} 
		\begin{tabular}{p{1.8cm}<{\centering}|p{1cm}<{\centering}|p{1cm}<{\centering}|p{1.3cm}<{\centering}|p{1.2cm}<{\centering}|p{1.2cm}<{\centering}}
			\hline
			&	\textsc{Deep\-Name}\xspace & $w/o$ Internal & $w/o$ Interaction & $w/o$ Siblings & $w/o$ Enclosing \\
			\hline
			Precision & 73.6\% & 52.8\% & 68.5\% & 70.4\% & 71.8\% \\
			Recall & 71.9\% & 49.3\% & 65.7\% & 68.2\% & 70.2\% \\
			F-score & 72.7\% & 51.0\% & 67.1\% & 69.3\% & 71.0\% \\
			Exact Match & 44.3\% & 24.3\% & 39.7\% & 40.5\% & 42.8\% \\
			\hline
		\end{tabular}
		%	\\
		%		INC: Internal Context; ITC: Interaction Context; SC: Sibling Context; EC: Enclosing Context
		\label{RQ3:2}
	\end{center}
	\vspace{-6pt}
	
\end{table}
\fi

\vspace{0.04in}
\noindent {\bf B. Impact of Weight Learning for Different Contexts.}
{\em The nature and the length of the sequences of sub-tokens in each
context might contribute differently}.
%
%As in Formula~\ref{eq:10}, the weight for each context is learned via
%$W_i$. This is a treatment to help the model learn the importance of
%different contexts.
To help our~model learn the importance of each context, we use weight
learning with $W_i$ in Formula~\ref{eq:10}.
%An alternative is to use the same equal weight for every context.
%We also conducted a study to learn the effectiveness of our weight
%learning.
We compared our model with the one having equal weights. The result
shows that with weight learning, {\tool} improves 5.13\%
in \code{accuracy} for consistency checking and 2.2\%
in \code{F-score} for name suggestion. Thus, our weight learning for
contexts positively contributes to accuracy.

\vspace{0.03in}
\subsubsection{\bf RQ4. Impact Analysis of {\em Copy} and {\em Non-Copy} Mechanisms}

\begin{table}[t]
	\caption{RQ4. {\em Copy/Non-Copy} in Consistency Checking}
	\vspace{-9pt}
        \small
        \tabcolsep 2pt
	\begin{center}
		\renewcommand{\arraystretch}{1}
		\begin{tabular}{p{0.5cm}<{\centering}|p{1.2cm}<{\centering}|p{1.2cm}<{\centering}|p{1.9cm}<{\centering}|p{3.4cm}<{\centering}}
			\hline
			&	& \code{Seq2seq} & \code{Seq2seq} & \code{Seq2seq+Copy+Non-copy} \\
                        & & & +\code{Copy} & (={\tool}) \\
			\hline
			\multirow{3}{*}{C} & Precision  & 64.7\% & 69.8\% & 72.3\% \\
			& Recall & 89.5\% & 91.2\% & 92.1\%  \\
			& F-score  & 75.1\% & 79.1\% & 81.0\% \\
			\hline
			\multirow{3}{*}{IC} & Precision  & 57.3\% & 59.6\% & 64.8\% \\
			& Recall  & 82.4\% & 83.1\% & 86.4\%\\
			& F-score  & 67.6\% & 72.1\% & 74.1\%\\
			\hline
			\multicolumn{2}{c|}{Accuracy}  & 68.8\% & 73.5\% & 75.8\% \\
			\hline
		\end{tabular}
		\label{RQ4:1}
	\end{center}
%	\vspace{-8pt}
\end{table}

In this experiment, we removed from our model both {\em Copy} and {\em
Non-Copy} mechanisms and used it as a baseline. We then added each
mechanism one-by-one to the baseline. As seen in
Tables~\ref{RQ4:1}--\ref{RQ4:2}, {\em Copy} mechanism helps improve
over the baseline model 6.8\% relatively in \code{accuracy} for
consistency checking, and 3.4\% relatively in \code{F-score} in name
suggestion.

The newly developed {\em Non-Copy} mechanism helps additionally
improve 3.1\% relatively over \code{seq2seq+Copy} in consistency
checking accuracy, and helps improve 2.7\% relatively in name
suggestion. Thus, {\em Non-copy} complements to {\em Copy} mechanism.

%Both mechanisms help improve the ranking of correct names in the
%resulting list. We will explain via an example later.

%As seen in Table~\ref{RQ4:1}, for method name consistency checking, copying and non-copying mechanism contribute to the overall performance of \tool. Without them, {\tool} performs worse in every metric. Specifically, the copying mechanism has the largest impact on \tool. By adding it, \tool improved 5.3\%, 6.7\% and 6.8\% on two f-score and accuracy. The Non-copying Mechanism also has a large impact on \tool, by using non-copy, \tool improved 2.4\%, 2.7\% and 3.14\% on two f-score and accuracy.

%As seen in Table~\ref{RQ4:2}, for method name generation, copying and non-copying mechanism also influence the results a lot. For details, the copying mechanism has a big influence on \tool for about 3.5\% and 8.1\% on f-score and exact match. Also, the non-copying mechanism have contribution of 2.0\% and 4.0\% for f-score and exact match.

\begin{table}[t]
	\caption{RQ4. {\em Copy/Non-Copy} in Name Suggestion}
	\vspace{-9pt}
        \tabcolsep 2pt
        \small
	\begin{center}
		\renewcommand{\arraystretch}{1}
		\begin{tabular}{p{1.2cm}<{\centering}|p{1.1cm}<{\centering}|p{1.9cm}<{\centering}|p{3.4cm}<{\centering}}
			\hline
			& \code{Seq2seq} & \code{Seq2seq+Copy} & \code{Seq2seq+Copy+Non-copy} (=\tool) \\
			\hline
			ExMatch  & 39.4\% & 42.6\% & 44.3\% \\
			Precision  & 69.3\% & 72.2\% & 73.6\% \\
			Recall & 68.5\% & 70.4\% & 71.9\%  \\
			F-score  & 68.9\% & 71.3\% & 72.7\% \\
			
			\hline
		\end{tabular}
		\label{RQ4:2}
	\end{center}
	\vspace{-10pt}
\end{table}

%Copying mechanism contributes to the \tool overall performance because it can directly copy existing key sub-tokens from the context. For example, in figure \ref{fig:motiv_2}, the key sub-tokens \code{get}, \code{Preferred} and \code{Size} all can be directly copied from interaction context while without copying mechanism, seq2seq model can only predict the sub-tokens that is hard to pick up these three key sub-tokens and predict them in the correct position. Non-copy mechanism also contribute to the performance too because, it can help reduce the search space of copying mechanism. For example, also in figure \ref{fig:motiv_2}, the sub-token \code{Size} followed sub-token \code{Preferred} in the training data for a few times, but some other sub-tokens like \code{Parent} never followed the sub-token \code{Preferred}. In this way, the non-copy mechanism will reduce the possibility of copying the sub-token \code{Parent} after the sub-token \code{Preferred}. So the sub-token \code{Size} will have higher possibility to be picked by the copying mechanism and it is how non-copying mechanism help on the overall performance.

%\input{sections/rq5-results}
\vspace{0.03in}
\subsubsection{{\bf Illustration}}

\begin{figure*}[ht]
	\centering
	\includegraphics[scale=0.26]{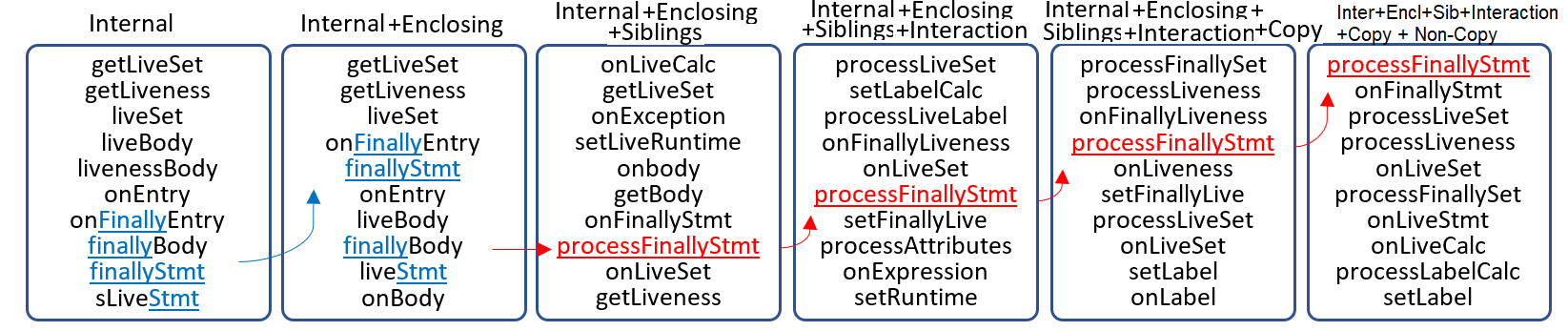}
	\caption{Method Name Recommendation Results (Top-10 Ranked List) for Fig.~\ref{fig:illustration}}
	\label{toplist}
	\vspace{-10pt}
\end{figure*}

\begin{figure}[t]
	\centering
	\renewcommand{\lstlistingname}{Method}
	\lstset{
		numbers=left,
		numberstyle= \tiny,
		keywordstyle= \color{blue!70},
		commentstyle= \color{red!50!green!50!blue!50},
		frame=shadowbox,
		rulesepcolor= \color{red!20!green!20!blue!20} ,
%
%xleftmargin=2em,xrightmargin=1em, aboveskip=1em,
%framexleftmargin=1.5em,
%language=Java,
%basicstyle=\scriptsize\ttfamily,
                xleftmargin=1.5em,xrightmargin=0em, aboveskip=1em,
		framexleftmargin=1.5em,
                numbersep= 5pt,
		language=Java,
                basicstyle=\scriptsize\ttfamily,
                numberstyle=\scriptsize\ttfamily,
                emphstyle=\bfseries,
		moredelim=**[is][\color{red}]{@}{@},
                escapeinside={(*}{*)}
	}
	\begin{lstlisting}
private static LiveCalc (*\fbox{XXXXXXXXXXX}*) (FinallyStmt s, LiveSet onEntry) {
	//Name: processFinallyStmt
	return liveness(s.getBody(), onEntry);
}
	\end{lstlisting}
        \vspace{-0.12in}
	\caption{A Correctly Suggested Method Name}
	\label{fig:illustration}
\end{figure}

Let us take an example in our experiment to illustrate the
effectiveness of each component in {\tool}. Fig.~\ref{toplist} shows
the top-ranked results for the name of the method given in
Fig.~\ref{fig:illustration} whose actual name
is \code{processFinallyStmt}. We show the top-ranked resulting lists
of the name for several variations of {\tool} in which we gradually
added each component/context to the previous model. The leftmost
column is the result of the model using only internal context (body
and interface).
%
%Let us use an example (Fig.~\ref{fig:illustration}) to illustrate the
%effectiveness of each component in {\tool}. Fig.~\ref{toplist} shows
%the top-ranked resulting lists of the method name for several variations of
%{\tool} in which we gradually added each component/context to the
%previous model. The first column is the result of the model using only
%internal context (body and interface).
%
The next column is the result from a new variation with the addition
of a new component. For example, the second column is the result from
a model that considers both internal and enclosing contexts; the
third one is from the model with internal, enclosing, and sibling
contexts, etc.

%The method at line 1 in Fig.~\ref{fig:illustration} is
%named \code{processFinallyStmt}.

Let us explain the resulting lists from all the variants and the
impacts of contributing components.
As seen, both the models \code{Internal}
%, which uses only the internal context, and
and \code{Internal+Enclosing}
%, which uses the internal context and enclosing context,
cannot suggest the correct name in the top-10 lists. The body and
interface contain the sub-tokens \code{Finally} and \code{Stmt}, but
do not contain the sub-token \code{process}. By adding the enclosing
context can only help improve the ranking of the candidate method
names that include sub-token \code{Finally} and \code{Stmt}.
However,
%the model $C$ with the internal, enclosing, and sibling
%contexts
by adding the sibling context, \code{Internal+Enclosing+Siblings} is able
to rank the correct method name
\code{processFinallyStmt} at the 8th position. The reason is that the
enclosing class in this case has several sibling methods with the
names starting with \code{process}, e.g., \code{processOperation},
\code{processLabeledStmtWrapper}, etc. Thus, \code{Internal+Enclosing+Siblings} is able to
learn the sub-token \code{process} from the sibling methods, and ranks
the correct name higher.

%However, when additionally considering the interaction context, the
%new model $D$
By adding the interaction
context, \code{Internal+Enclosing+Sibl\-ings+Interaction} can
rank the correct name at the 6th~position. The reason is that the
body and interface of the callee method \code{liveness} contain
the occurrences of \code{process} and \code{stmts}.

With the addition of the {\em Copy} mechanism, the new
model \code{Internal}+\code{Enclosing}+\code{Siblings}+\code{Interaction}+\code{Copy}
improves the ranking of the correct name to the 4th position.  The
reason is that {\em Copy} mechanism can emphasize on the copying of
the popular and important sub-tokens, e.g., \code{process},
\code{liveness}. As seen, the names with the sub-tokens \code{process}
and \code{liveness} are ranked at the top 5 positions in the column
corresponding to this model.

Finally, {\tool}, with the addition of {\em Non-copy} mechanism, is
able to rank the correct name at the top position. The reason~is that
{\em Non-copy} can learn that {\em the sub-token~\code{Set}
must not follow \code{processFinally}}. Thus, \code{processFinallySet}
is pushed down, and the sub-token \code{Stmt}
following \code{processFinally} to produce the correct name
\code{processFinallyStmt} is pushed to the top.

\subsubsection{{\bf RQ5. Accuracy Analysis on Method Name Suggestion}}

\begin{figure}[t]
	\centering
	\includegraphics[width = 2.34in]{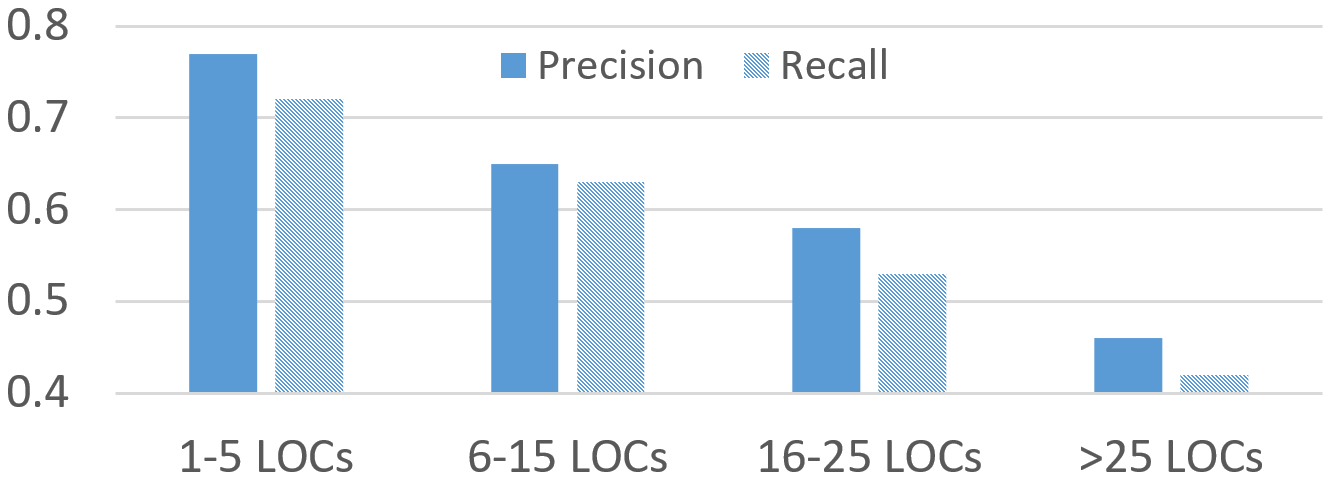}
%	\vspace{-3pt}
	\caption{Accuracy by different Methods' Sizes in Test Set}
	\label{RQ5}
%	\vspace{-10pt}
\end{figure}

We study the results on the suggested method names that {\em were
not in the training data}. There are +173K (11.9\%) out of +1,445K
generated method name that were un-seen during the
training. The \code{Precision}, \code{Recall}, and \code{F-score} for this
set are 57.6\%, 55.1\%, and 56.3\%, respectively. Importantly, in
17.4\% of these generated cases (i.e., 2.1\% total cases), the
generated names exactly match the expected names in the oracle. These numbers
show that {\tool} performs well for the un-seen method names and really
learns to suggest natural names, rather than retrieving method names
that have been stored in the training dataset.

%the exact match percentage is 17.4\% which prove that \tool can do
%well on un-seen method name which means \tool really learns and
%predicts method names instead of retrieving existing method names from
%training dataset.

\vspace{0.03in}
\noindent {\bf Accuracy by the Sizes of Methods in Test Set.}
As seen in Fig.~\ref{RQ5}, \tool works well on the methods with the
regular sizes of 1-25 LOCs. Even with the longer methods (+25 LOCs),
\code{Precision} and \code{Recall} decrease gracefully at 46.3\% and 41.9\%,
respectively. This shows that predicting the name becomes harder for
longer methods. Even so, in +8K cases of 33K long methods with +25
LOCs, {\tool} produced the exact-match names with the correct ones.

%$And if the method length increase, our
%approach accuracy decreased, but the recall and precision are still
%have 46.3\% and 41.9\%. Also, the figure also proves that when the
%method is getting longer, our approach still can extract useful
%features from interaction context and siblings to generate method
%names in the correct way which shows that considering different types
%of context is very useful.

\vspace{0.03in}
\subsubsection{{\bf RQ6. Live Study}}

\begin{table}[t]
	\caption{RQ7. Pull Requests of Real-world Projects}
%	\vspace{-7pt}
        \small
	\begin{center}
		\renewcommand{\arraystretch}{1}
		\begin{tabular}{p{1cm}<{\centering}|p{1cm}<{\centering}|p{1cm}<{\centering}|p{1.5cm}<{\centering}|p{1cm}<{\centering}}
			\hline
			%\multicolumn{2}{c|}{Accept} & \multicolumn{2}{c|}{Agree} & \multirow{2}{*}{Disagree} & \multirow{2}{*}{No Reply} & \multirow{2}{*}{Total}\\
		%	Merged & Approved & Not Fix & Cannot Fix & & & \\
		
			Accept & Agree & Disagree & No-reply & Total\\
			\hline
			 12 & 18 & 11 & 9 & 50 \\

			\hline
		\end{tabular}
		\label{RQ7}
	\end{center}
%	\vspace{-5pt}
\end{table}

To evaluate the usefulness of our tool, we conducted a study on 100
randomly chosen, active Java projects in GitHub. We used {\tool}
trained as in RQ1 to detect inconsistent method names, then submitted
pull requests (PRs) of method renaming suggested by the tool and
assessed PR acceptance rates.~Overall, it identified 3K out of 133K
methods as inconsistent.
%
%To avoid putting too much burden on the developers with too many PRs,
To avoid much work for developers, we randomly selected and made only
50~pull~requests.
%cases of inconsistent method names and suggested names.
%suggested by {\tool}.
%
%We performed method renaming refactoring in the projects
%and submitted the changes as pull requests.

%to repositories.
%We then prepared a patch by performing a method name renaming
%refactoring and compiling the entire corresponding project. We
%submitted the changes as a pull request to the project repository.

As seen in Table~\ref{RQ7}, among 50 PRs, 12 cases were approved and
merged by the development teams. Additionally, 18 PRs have been
validated and approved by the team members. For those cases, the teams
acknowledged that the current method names are misleading, and agreed
with the suggested names as providing more meaningful names. However,
at the time of writing, the PRs have not been merged into the main
branch due to the additional requirements of reviewing or testing.
%the PR changes have not been merged at the time of this writing since
%the teams require additional tasks, \eg reviewing and unit testing
%before PRs can be merged to the main branch of a project.
%The approved method names belong to a diverse group including public,
%private, or protected methods.
%Tien
In 11 cases, the developers disagreed with our suggested names. In
some cases, the suggested names do not conform to coding conventions
in the project. Some cases involve template code. In
some cases, the names are of the methods that override the external
libraries. There are still 9 cases that we did not get responses.
In brief, in 30/50 cases, the developers confirmed that the names
suggested by {\tool} are more meaningful than the current names.  This
shows that {\tool} is useful in real-world projects in both detecting
inconsistent method names and suggesting new names.

\subsubsection{Threats to Validity}
Our data has only Java code.
%However, for comparison, we used the same datasets and metrics as
%previous works.
For code2vec~\cite{alon_popl19}, we used the same metrics for
comparison (i.e., the accuracy for a set of method names are
the average of those for all names). We did not have a
statistical test in comparison since they did not provide individual
resulting names. Running their tool requires a high-computational machine. We could not run Liu {\em et
al.}~\cite{kim-icse19}'s name suggestion on our dataset
despite our efforts trying and contacting the authors without
responses.

\subsubsection{Limitations}

Despite the above successes, {\tool} also has the aspects that need to
be improved. As any other ML approaches, it has the out-of-vocabulary
issue. That is, it cannot generate a sub-token that has never been
seen in the training data. However, as shown in the empirical
evaluation section, {\tool} is able to generate a new method name from
the sub-tokens that it has encountered in the training dataset.
Because {\tool} does not analyze the entire project, it does not
perform well for the overriding methods, and the methods that override
the APIs in the external libraries. A potential solution is to
integrate our ML direction with program analysis to guide the process
of the method name generation. Moreover, {\tool} does not work well
for the method names with one single sub-token, or the methods with
long bodies or the long callers/callees.

%\section{Threats to Validity}

%\input{sections/discu}
%\vspace{-3pt}
\section{Related Work}
\label{related}

There are two categories of approaches for method name inconsistency
detection and suggestion. The first one is Information Retrieval
(IR). Liu {\em et al.}~\cite{kim-icse19} relies on the principle that
methods with similar bodies have similar names.  In our experiment, we
showed that such principle does not hold in many cases. Jiang {\em et
  al.}~\cite{jiang-ase19} searches for the methods having similar
return type and parameters, as well as heuristics,~to derive method
names. The key advantage of these IR-based approaches is that they are
light-weighted and do not require high computational power.  Their
key disadvantage is that because searching in the set of
already-existed names, they cannot generate a new name that were not
in the training data.

The second category is machine learning
(ML). MNire~\cite{icse20-methodname} explores the sub-tokens appearing in the methods' bodies and interfaces.
%It uses a seq2seq encoder decoder to predict the method name from the
%sequences of those sub-tokens.
Allamanis {\em et al.}~\cite{Allamanis_16} use a neural network with
attention and convolution to summarize code into descriptive
summaries. Allamanis {\em et al}~\cite{Allamanis_15} project all the
sub-tokens in the entities' names in the method bodies into the
same vector space and cluster them to compose method name. MNire has
been shown to outperform
\code{code2vec}~\cite{alon_popl19}, which outperforms Allamanis {\em
  et al.}~\cite{Allamanis_16} and Allamanis {\em et
  al}~\cite{Allamanis_15}. Those ML-based approaches all rely only
on the method's body and~interface.

There are several approaches for code
embeddings. Code2vec~\cite{alon_popl19} abstracts source code by the
paths over the AST to produce
vectors. Code2seq~\cite{alon2018code2seq} generates a word sequence
from the structure of source code. Ke and Su~\cite{ke-pldi20}'s
approach builds the embeddings to capture structures and semantics of
a program.
However, as shown in MNire~\cite{icse20-methodname}, using code
structure, AST, PDG is too strict in predicting method names.
%They showed that using sequences of sub-tokens achieves better
%accuracy than using code structure.

%They are code embedding techniques, not name suggestion ones.

%the sub-tokens are more important in predicting the method names than
%the code structures of the method's body. We also showed that with
%rich contexts, {\tool} achieves higher accuracy than code2vec.

%Name recovery
There are several approaches to {\em predict the names or types of
  program entities} within the method
bodies~\cite{icse18,JSNice2015,icse19,JSNaughty2017}. While
JSNeat~\cite{icse19} searches for names in a large corpus to recover
variable names in minified code, JSNice~\cite{JSNice2015} and
JSNaughty~\cite{JSNaughty2017} use CRF and machine
translation. Naturalize~\cite{barr-codeconvention-fse14} learns and
enforces a consistent naming conventions.

Several approaches use ML to generate texts from
code~\cite{h16,hu-icpc18,iyer16,david-ase19,wan18} and vice
versa~\cite{gu-fse16,anycode-oopsla15,icsme18,Raghothaman-ICSE16}, or
code migration~\cite{icsme16,ase14,fse13,ase15}. Zheng {\em et al.}~\cite{zheng-fcs18} uses
AST structure for such statistical machine translation to produce
comments. CODE-NN~\cite{iyer16} uses LSTM on code sequence to model
the conditional distribution of a summary to produce word by word.
DeepCom~\cite{hu-icpc18} has a traversal on AST for
flattening, and uses seq2seq to produce code summary. Wan {\em
  et al.}~\cite{wan18} use a deep reinforcement learning on AST and code sequence.
%
%Statistical NLP was used to generate code from text, e.g.,
%SWIM~\cite{Raghothaman-ICSE16}, DeepAPI~\cite{gu-fse16},
%Anycode~\cite{anycode-oopsla15}, etc.
There are several studies on name consistency and naming
convention~\cite{Venera-emse16,binkley-msr11,debug09,kim-emse16}.

\section{Conclusion}

We introduce {\tool}, a context-based deep learning approach for
inconsistency checking and method name suggestion. The
following key ideas enable our approach (1) characterizing a method by
the surrounding methods that are interaction or siblings method for
the method we are studying; (2) learning the representation for the
method with multiple contexts; (3) using sub-token copying and
non-copying mechanisms to help better predict the name. We conducted
several experiments to evaluate {\tool}.

Our results showed high accuracy and usefulness of {\tool} in
real-world projects.
%We conducted several experiments to evaluate {\tool} on large datasets
%with +14M methods.
For consistency checking, {\tool} improves the state-of-the-art
approach by 2.1\%, 19.6\%, and 11.9\% relatively in recall, precision,
and F-score, respectively. For name suggestion, {\tool} improves
relatively over the existing approaches in precision (1.8\%--30.5\%),
recall (8.8\%--46.1\%), and F-score (5.2\%--38.2\%). In the assessment
of {\tool}'s usefulness in real-world projects, the team members agree
that our suggested method names are more meaningful than the current
names in 30/50 cases. For future work, we plan to integrate our ML
direction with program analysis to improve both accuracy
and efficiency.

%To assess {\tool}'s usefulness, we detected inconsistent methods and
%suggested new method names in active projects.  Among 50 pull
%requests, 12 were merged into the main branch. In total, in 30/50
%cases, the team members agree that our suggested method names are more
%meaningful than the current names.

%We have processed several empirical studies to evaluate \tool on method consistency checking and method name recommendation tasks. \tool is able to outperform the existing state-of-the-art consistency checking approaches and method name recommendation approaches. Our experiment results shows that \tool can have a relevant improvement by (14.4\%--33.9\%) and (15.4\%--33.6\%) on F-score and accuracy when comparing with consistency checking baselines. And also can improve the F-score and exact match percentage for (11.6\%--29.8\%) and 13.9\% when comparing with method name recommendation baselines. And also, our live study shows that 30 of 50 names suggested by \tool are more meaningful than original ones that proves \tool is also useful for real program developing processes.

\section*{Acknowledgment}
This work was supported in part by the US National Science
Foundation (NSF) grants CCF-1723215, CCF-1723432, TWC-1723198,
CCF-1518897, and CNS-1513263.

%\newpage

\balance

\bibliographystyle{IEEETrans}
\bibliography{sections/ref-1,sections/ref-2,sections/ref-3,sections/ref-4,sections/ref-5}

\end{document}